\documentclass[journal,twoside]{IEEEtran}

\usepackage{color}
\usepackage{bbm}
\usepackage{amsmath,amsfonts}
\usepackage{dsfont}
\usepackage{graphicx}
\usepackage[hang]{subfigure}
\usepackage{citesort}
\usepackage{mathtools}

\usepackage{ntheorem}
\theoremstyle{plain}
\theoremheaderfont{\itshape}\theorembodyfont{\itshape}
\theoremseparator{.}
\newtheorem{theorem}{Theorem}
\newtheorem{corollary}{Corollary}
\newtheorem{proposition}{Proposition}
\newtheorem{lemma}{Lemma}

\theoremheaderfont{\itshape}\theorembodyfont{\upshape}
\theoremseparator{.}
\newtheorem{definition}{Definition}
\newtheorem{remark}{Remark}

\newcommand{\sA}{\mathcal{A}}
\newcommand{\sC}{\mathcal{C}}
\newcommand{\sD}{\mathcal{D}}

\newcommand{\sM}{\mathcal{M}}
\newcommand{\sP}{\mathcal{P}}
\newcommand{\sS}{\mathcal{S}}

\newcommand{\sU}{\mathcal{U}}
\newcommand{\sX}{\mathcal{X}}	
\newcommand{\sY}{\mathcal{Y}}
\newcommand{\sZ}{\mathcal{Z}}
\newcommand{\N}{\mathbb{N}}
\newcommand{\R}{\mathbb{R}}
\newcommand{\C}{\mathbb{C}}
\newcommand{\E}{\mathbb{E}}
\newcommand{\tr}{\mathop{\mathrm{tr}}\nolimits}
\newcommand{\diag}{\mathop{\mathrm{diag}}\nolimits}
\newcommand{\ba}{{\bf a}}
\newcommand{\bA}{{\bf A}}
\newcommand{\bB}{{\bf B}}
\newcommand{\bW}{{\bf W}}
\newcommand{\bR}{{\bf R}}
\newcommand{\bH}{{\bf H}}
\newcommand{\bx}{{\bf x}}
\newcommand{\by}{{\bf y}}
\newcommand{\bxi}{{\boldsymbol \xi}}
\newcommand{\bI}{{\bf I}}

\newcommand{\bU}{{\bf U}}
\newcommand{\bV}{{\bf V}}
\newcommand{\bSg}{{\boldsymbol \Sigma}}
\newcommand{\bL}{{\boldsymbol \Lambda}}
\newcommand{\comp}{\mathfrak{W}}
\renewcommand{\vec}[1]{\boldsymbol{#1}}
\newcommand{\bsS}{\vec{\sS}}
\newcommand{\sSo}{\vec{\sS}_1}
\newcommand{\sSt}{\vec{\sS}_2}
\newcommand{\Wtil}{\overline{W}}
\newcommand{\Ptil}{\overline{P}}
\def\bal#1\eal{\begin{align}#1\end{align}}

\begin{document}

\title{The Secrecy Capacity of Compound Gaussian MIMO Wiretap Channels}
\author{Rafael F. Schaefer,~\IEEEmembership{Member,~IEEE}, and Sergey Loyka
\thanks{This work was supported by the German Research Foundation (DFG) under Grant WY 151/2-1. This paper was presented in part at the IEEE Information Theory Workshop in 2013 and the 52nd Annual Allerton Conference on Communication, Control, and Computing in 2014.}
\thanks{R. F. Schaefer is with the Department of Electrical Engineering, Princeton University, Princeton, NJ 08544, USA (e-mail: rafaelfs@princeton.edu).}
\thanks{S. Loyka is with the School of Electrical Engineering and Computer Science, University of Ottawa, Ottawa, ON K1N 6N5, Canada (e-mail: sergey.loyka@uottawa.ca).}
}
\IEEEoverridecommandlockouts
\maketitle

\markboth{IEEE Transactions on Information Theory}{Schaefer and Loyka: Secrecy Capacity of Compound Gaussian MIMO Wiretap Channels}

\begin{abstract}
Strong secrecy capacity of compound wiretap channels is studied. The known lower bounds for the secrecy capacity of compound finite-state memoryless channels under discrete alphabets are extended to arbitrary uncertainty sets and continuous alphabets under the strong secrecy criterion. The conditions under which these bounds are tight are given. Under the saddle-point condition, the compound secrecy capacity is shown to be equal to that of the worst-case channel. Based on this, the compound Gaussian MIMO wiretap channel is studied under the spectral norm constraint and without the degradedness assumption. First, it is assumed that only the eavesdropper channel is unknown, but is known to have a bounded spectral norm (maximum channel gain). The compound secrecy capacity is established  in a closed form and the optimal signaling is identified: the compound capacity equals the worst-case channel capacity thus establishing the saddle-point property; the optimal signaling is Gaussian and on the eigenvectors of the legitimate channel and the worst-case eavesdropper is isotropic. The eigenmode power allocation somewhat resembles the standard water-filling but is not identical to it. More general uncertainty sets are considered and the existence of a maximum element is shown to be sufficient for a saddle-point to exist, so that signaling on the worst-case channel achieves the compound capacity of the whole class of channels. The case of rank-constrained eavesdropper is considered and the respective compound secrecy capacity is established. Subsequently, the case of additive uncertainty in the legitimate channel, in addition to the unknown eavesdropper channel,  is studied. Its compound secrecy capacity and the optimal signaling are established in a closed-form as well, revealing the same saddle-point property. When a saddle-point exists under strong secrecy, strong and weak secrecy compound capacities are equal.
\end{abstract}


\section{Introduction}
\label{sec:introduction}

The nature of the wireless medium makes wireless communication systems inherently vulnerable for eavesdropping. In this context, the concept of information theoretic security is instrumental since it solely uses the physical properties of the wireless channel in order to establish security. Information theoretic security was initiated by Shannon \cite{Shannon49CommunicationTheorySecrecySystems} and studied later by Wyner, who introduced the now-popular \emph{wiretap channel} \cite{Wyner75WiretapChannel} modeling the simplest scenario involving security with one legitimate transmitter-receiver pair and one wiretapper (eavesdropper) to be kept secret. There is currently a growing interest in information theoretic security, see e.g.  \cite{Liang09InformationTheoreticSecurity,Jorswieck10SecrecyPhysicalLayer,Liu10SecuringWirelessCommunications,Bloch11InformationTheoreticSecrecy}.

Since spatial multiple-input multiple-output (MIMO) techniques can improve the performance significantly \cite{BiglieriG07MIMOWirelessCommunications}, MIMO architectures have been identified as indispensable for future wireless systems. Accordingly, investigation of information theoretic security for MIMO systems is becoming more and more attractive. The secrecy capacity of the Gaussian MIMO wiretap channel is established in \cite{KhistiWornell10MIMOWiretap1,KhistiWornell10MIMOWiretap2,Oggier11MIMOWiretap,Liu09MIMOWiretapSecrecy} under full channel state information (CSI), where it turns out that Gaussian signaling is optimal. Subsequently, the optimal transmit covariance matrix has then been found under the matrix power constraint in \cite{Bustin09MMSEMIMOWiretap} and under the total power constraint for a number of special cases \cite{KhistiWornell10MIMOWiretap2,KhistiWornell10MIMOWiretap1,Loyka12OptimalSignalingMIMOWiretap,LiXXMIMOWiretap}.

Due to the dynamic nature of the wireless medium, but also due to implementation issues, practical systems always suffer from channel uncertainty and estimation/feedback inaccuracy. Thus, the provision of accurate channel state information to the transmitter is a major challenge for wireless communication systems. Along with this, it is hardly possible to expect that the eavesdropper will share its channel with the transmitter to make the eavesdropping harder, which makes the perfect eavesdropper CSI model more than questionable. A reasonable and well-accepted approach to this problem is to assume that the exact channel realization is not known; it is only known that it remains fixed during the entire transmission and that it belongs to a known set of channels (uncertainty set), which results in the concept of \emph{compound channels} \cite{Blackwell59Compound,Wolfowitz60SimultaneousChannels}.

The discrete memoryless compound wiretap channel with a countably-finite uncertainty set (i.e. finite-state channels) is studied in \cite{Liang09CompoundWiretapChannels,Bjelakovic13CompoundWiretap}. Its secrecy capacity is established under the degradedness assumption, where all possible realizations of the eavesdropper channel must be degraded with respect to all possible realizations of the legitimate channel. When this condition is not satisfied, only an achievable secrecy rate is given while the secrecy capacity for the general case remains unknown.

The corresponding compound Gaussian MIMO wiretap channel with countably-finite uncertainty sets is analyzed in \cite{Liang09CompoundWiretapChannels}. Similarly to the discrete memoryless case, its secrecy capacity is established, again, only under the degradedness assumption. When the channel is not degraded, the secrecy capacity itself remains unknown and only an achievable secrecy rate is obtained. In \cite{Ekrem10MIMOCompoundWiretap}, the special case of compound wiretap channels with two possible channel states for the legitimate receiver and known eavesdropper channel is studied. Its secrecy capacity is established under the degradedness assumption and an achievable rate is given in the general (non-degraded) case while the capacity is unknown. Interference alignment for the compound Gaussian MIMO wiretap channel is explored in \cite{Khisti11InterferenceAlignmentMIMOCompoundWiretap}. A Gaussian MIMO wiretap channel where the noiseless eavesdropper channel is arbitrarily varying is considered in \cite{HeYener10MIMOWiretap}. Its achievable secrecy rate (i.e. lower bound to the secrecy capacity) is given and the secrecy degrees of freedom are established, while its capacity remains unknown. Since degrees of freedom require $\text{SNR}\rightarrow\infty$, it is not clear what finite-SNR implications are\footnote{Two systems having the same degrees of freedom may have vastly-different capacities, even at high SNR, see e.g. \cite{Loyka10FiniteSNRDMTLargeMIMO}.}.

The discrete memoryless compound broadcast channel with confidential messages is studied in \cite{Schaefer14CompoundBCC} and its strong secrecy capacity region is established in a multi-letter form. The corresponding Gaussian MIMO broadcast channel is considered in \cite{Kobayashi09CompoundMIMOBCC} and its achievable degree-of-freedom region is established, but not the capacity region itself.

In all the previous studies, the compound secrecy capacity has been established only for the special case of degraded channels with countably-finite uncertainty sets \cite{Liang09CompoundWiretapChannels,Bjelakovic13CompoundWiretap,Ekrem10MIMOCompoundWiretap}. Accordingly, it is not clear if these results hold for more general (e.g. uncountable) or arbitrary uncertainty sets as well or how these results extend to non-degraded channels. For such non-degraded channels, only an achievable secrecy rate is obtained with the consequence that it is not clear how far away this rate is from its actual capacity. The achievable secrecy rate is studied from a worst-case secrecy rate maximization point of view in \cite{Wolf15WorstCaseSecrecyRatesMIMOME}. Another approach is taken in \cite{Khisti11InterferenceAlignmentMIMOCompoundWiretap,HeYener10MIMOWiretap} by studying the behavior for $\text{SNR}\rightarrow\infty$. However, this does not provide any insights on the secrecy capacity or its behavior for the practical relevant case of finite $\text{SNR}<\infty$.

In this paper, we address all these limitations and establish the (strong) secrecy capacity of compound Gaussian MIMO channels for a broad class of uncertainty sets (not only finite or countable) and without the degradedness assumption.
We make use of the compound wiretap model, where the legitimate channel is perfectly known and the eavesdropper channel is not known to the transmitter but is known to have a bounded spectral norm (maximum channel gain), both being fixed during the whole transmission duration. This represents a quasi-static scenario where the eavesdropper cannot approach the transmitter closer than a certain protection distance so that its channel gain is bounded (due to the propagation path loss) but is unconstrained otherwise. This automatically implies only a minimal eavesdropper CSI at the transmitter, which reflects well the natural eavesdropper desire to be confidential and its lack of cooperation. Throughout the paper, full CSI at the eavesdropper is assumed (the safest assumption from the secrecy perspective). We make no assumptions of degradedness. The eavesdropper channel uncertainty scenario is further extended to the case where the legitimate channel is also allowed to have (additive) uncertainty, which represents channel estimation and feedback link limitations, and to the case of more general eavesdropper uncertainty sets, which may be non-isotropic.

The compound secrecy capacity is established in two main steps. First, we consider the corresponding discrete memoryless (DMC) channel in Section \ref{sec:dmc}. For this channel model, an achievable (strong) secrecy rate was obtained in \cite{Bjelakovic13CompoundWiretap} for countably-finite uncertainty sets. Building on this result, we establish a lower bound for the compound (strong) secrecy capacity under arbitrary uncertainty sets (not necessarily finite or countable) in Theorem \ref{the:arbitrary}, which is subsequently extended to continuous alphabets in Theorem~\ref{the:continuous} using the set partitioning (quantization) arguments adopted to compound channels in~\cite{Mitran06CompoundSideInformation}. The conditions under which these bounds are tight are given, thus establishing the secrecy capacity. Under the saddle-point condition, the compound secrecy capacity is shown to be equal to that of the worst-case channel (so that any code designed for the worst-case channel also works on the entire class of channels in the uncertainty set).

Secondly, the (strong) secrecy capacity of the compound Gaussian MIMO channel is established in Theorem \ref{thm.M1} for the eavesdropper uncertainty with bounded spectral norm and without the degradedness assumption. This is done by establishing first an achievable rate of this channel in Corollary \ref{cor:mimo}\footnote{Unlike \cite{Liang09CompoundWiretapChannels}, this is done for arbitrary (compact) uncertainty sets, not just countable or finite, and under the strong secrecy constraint.}. Then, in Section \ref{sec:uncert}, the worst-case secrecy capacity (i.e. the capacity of the worst-case channel in the set) is obtained and the saddle-point property is established in the form $\max\min = \min\max$, where the maximization is over the transmit covariance and minimization is over the eavesdropper channel uncertainty. The saddle-point property has the well-known game-theoretic interpretation: the mini-max zero-sum game is between the transmitter (who controls the transmitted signal distribution) and the eavesdropper (who controls the channel); neither player can deviate from an optimal strategy without incurring penalty provided the other player follows it.

Combining all these, we establish the secrecy capacity of the compound Gaussian MIMO channel in a closed-form, which also equals the worst-case capacity, so that a code designed for the worst-case channel works over the whole class of channels as well. The optimal signaling is Gaussian and on the eigenvectors of the legitimate channel, with power allocation somewhat similar but not identical to the regular water-filling. The worst-case eavesdropper is isotropic with the maximum allowed channel gain. This result is then extended to a broader class of compound channels, where the uncertainty set is only required to have a dominant (maximum/maximal) element and may be non-isotropic. It is shown that the existence of a maximum element in the eavesdropper uncertainty set is sufficient for a saddle-point to exist, so that the compound capacity equals the worst-case one and signaling on the worst-case channel achieves the capacity of the whole class of channels. The high/low SNR regimes are considered and the condition for beamforming optimality is given. When the eavesdropper uncertainty is sufficiently large, beamforming is optimal at any SNR. The case of rank-constrained eavesdropper is considered, motivated by the scenario where the transmitter is a base station with a large number of antennas while the receiver/eavesdropper are handsets with a small number of antennas. Under this non-convex constraint (in addition to the convex spectral norm constraint), there is no maximum element in the uncertainty set, yet the saddle-point property is shown to hold and the compound secrecy capacity is established. Subsequently, a more general case of two-sided channel uncertainty is studied in Section \ref{sec:double-sided}, where the legitimate channel is also allowed to have (additive) uncertainty. This reflects the assumption that the legitimate receiver will share its CSI with the transmitter, but limitations in feedback link and channel estimation result in channel uncertainty. The corresponding compound secrecy capacity is established and shown to be equal to the secrecy capacity of the worst-case channel in the uncertainty set, so that the saddle-point property still holds. The optimal signaling is still on the eigenmodes of the legitimate channel and the worst-case eavesdropper is isotropic.

While it was established in \cite{Csiszar96SecrecyCapacity,Maurer00WeakToStrongSecrecy} that the strong and weak secrecy capacities are the same for regular (non-compound or known) channels, Section \ref{sec:weak vs. strong} demonstrates that the same holds for compound channels if the saddle-point property holds under strong secrecy.

Finally, Section \ref{sec:conclusion} concludes the paper.

\emph{Notations:} Discrete random variables are denoted by capital letters and their realizations and ranges by lower case and script letters, respectively; scalars, vectors, and matrices are denoted by lower case letters, bold lower case letters, and bold capital letters; $\N$, $\R_+$, and $\C$ are the sets of positive integers, non-negative real, and complex numbers respectively; $|\sA|$ and $\sA^c$ denote the cardinality and the complement of the set $\sA$. $I(\cdot;\cdot)$ is the mutual information and $H_2(\cdot)$ is the binary entropy function; $\sP(\cdot)$ is the set of all probability distributions and $\E\{\cdot\}$ is statistical expectation; $X \rightarrow Y \rightarrow Z$ denotes a Markov chain of random variables $X$, $Y$, and $Z$ in this order; $\bA^T$, $\bA^+$, and $|\bA|$ are the transposition, Hermitian conjugation, and determinant of $\bA$; $\tr\bA$ is the trace of the matrix $\bA$ and $\diag(\ba)$ is a diagonal matrix with elements given by $\ba$; $\bA\geq\bB$ means the matrix $\bA-\bB$ is positive semi-definite; $\bI$ is the identity matrix.

\section{Discrete Memoryless Channels}
\label{sec:dmc}

In this section we consider discrete memoryless channels (DMCs) with finite input and output alphabets. Building on earlier results in \cite{Bjelakovic13CompoundWiretap} for finite-state channels, an achievable secrecy rate is established for the general case of arbitrary uncertainty sets (not limited to finite or countable), which is subsequently extended to continuous alphabets in Section \ref{sec:continuous}.

\subsection{Compound Wiretap Channel}
\label{sec:dmc_wiretap}

Let $\sX$ and $\sY$, $\sZ$ be countably-finite input and output sets and $\sS$ be a set which will model the channel uncertainty. The channels to the legitimate receiver and the eavesdropper (wiretapper) are given by $W_s:\sX\times\sS\rightarrow\sP(\sY)$ and $V_s:\sX\times\sS\rightarrow\sP(\sZ)$, respectively, where $s\in\sS$ is a channel state. For a fixed state $s\in\sS$, input and output sequences $x^n\in\sX^n$ and $y^n\in\sY^n$, $z^n\in\sZ^n$ of block length $n$, the discrete memoryless channels are given by $W_s^n(y^n|x^n)=\prod_{i=1}^nW_s(y_i|x_i)$ and $V_s^n(z^n|x^n)=\prod_{i=1}^nV_s(z_i|x_i)$. The channels are assume to be quay-static: $s$ is selected at the beginning and is held constant during the entire transmission.

\begin{definition}
\label{def:compound}
The discrete memoryless \emph{compound wiretap channel} $\mathfrak{W}$ is given by
\begin{equation*}
	\comp=\big\{(W_s,V_s):s\in\sS\big\}.
\end{equation*}
\end{definition}

\begin{remark}
\label{rem:compound}
This includes the widely adopted model of the form $\comp=\{(W_{s_1},V_{s_2}):s_1\in\sS_1,s_2\in\sS_2\}$ with $\sS_1\neq\sS_2$ as one can always construct a new set of the form $\sS=\sS_1\times\sS_2$.
\end{remark}

\begin{definition}
\label{def:code}
An $(n,M_n)$-\emph{code} $\sC_n$ for the compound wiretap channel consists of a stochastic encoder at the transmitter
\begin{equation}
	E:\sM_n\rightarrow\sP(\sX^n),
	\label{eq:dmc_encoder}
\end{equation}
i.e., a stochastic matrix, with a set of messages $\sM_n=\{1,...,M_n\}$ and a decoder at the legitimate receiver described by a collection of disjoint decoding sets
\begin{equation}
	\{\sD_m\subset\sY^n:m\in\sM_n \big\},
	\label{eq:dmc_decoder}
\end{equation}
so that $\widehat{m}=m$ if $y^n \in \sD_m$, where $\widehat{m}$ is the decoded message at the receiver.
\end{definition}

The encoder in \eqref{eq:dmc_encoder} is allowed to be stochastic (this in fact is essential for achieving secrecy) which means that it is specified by conditional probabilities $E(x^n|m)$ with $\sum_{x^n\in\sX^n}E(x^n|m)=1$  for each $m\in\sM_n$. Then, $E(x^n|m)$ denotes the probability that the message $m\in\sM_n$ is encoded as $x^n\in\sX^n$.

Then for an $(n,M_n)$-code $\sC_n$, the maximum probability of decoding error at the legitimate receiver is given by
\begin{equation}
	e_n = \sup_{s\in\sS}\max_{m\in\sM_n}\sum_{x^n\in\sX^n}W_s^n(\sD_m^c|x^n)E(x^n|m).
	\label{eq:dmc_error}
\end{equation}

\begin{remark}
\label{rem:csi}
Throughout the whole paper we assume that the transmitter and legitimate receiver do not have full CSI, i.e., they do not know the actual realization $s\in\sS$ but do know the uncertainty set $\sS$. Accordingly, encoder \eqref{eq:dmc_encoder} and decoder \eqref{eq:dmc_decoder} are universal and do not depend on the particular realization. On the other hand, we make a conservative (and safest from secrecy perspective) assumption that the  eavesdropper has perfect CSI of both channels (to the legitimate receiver and its own).
\end{remark}

To keep the transmitted message secret from the eavesdropper for all channel realizations $s\in\sS$, we require the information leaked to the eavesdropper to be arbitrarily small, i.e.
\begin{equation}
	\sup_{s\in\sS}I(M;Z_s^n)\leq\epsilon_n
	\label{eq:dmc_secrecy}
\end{equation}
for some $\epsilon_n>0$ and $\epsilon_n \rightarrow 0$ as $n \rightarrow\infty$, where $M$ is the random variable uniformly distributed over the set of messages $\sM_n$ and $Z_s^n=[Z_{s,1},Z_{s,2},...,Z_{s,n}]$ is the eavesdropper channel output for channel realization $s\in\sS$. This criterion is known as \emph{strong secrecy} \cite{Csiszar96SecrecyCapacity,Maurer00WeakToStrongSecrecy}.

\begin{remark}
\label{rem:strongsecrecy}
The vanishing information leakage to the eavesdropper implies that its bit error probability $P_b$ approaches $1/2$ as $n\rightarrow\infty$ (and thus codeword error probability approaches 1) and the speed of convergence depends on the secrecy criterion adopted. In particular, it can be shown (using Fano's inequality) that
\bal
P_b &= \frac{1}{2} - o(1)\ \mbox{under weak secrecy}, \\
P_b &= \frac{1}{2} - o\left(\frac{1}{\sqrt{n}}\right)\ \mbox{under strong secrecy},
\eal
so that $P_b \rightarrow 1/2$ in any case, but the speed of convergence can be arbitrarily slow under weak secrecy, while it is at least as $1/\sqrt{n}$ under strong secrecy. Using the recent result in \cite{Bjelakovic13CompoundWiretap} on exponential convergence of information leakage to zero for the discrete memoryless channel (see \eqref{eq:dmc_strong}), it can be further shown that
\bal
P_b = \frac{1}{2} - O(e^{-an})
\eal
i.e. exponentially fast in that scenario, where $a>0$. This provides an operational meaning for secrecy criteria.
\end{remark}

\begin{definition}
\label{def:achievable}
A non-negative number $R_s$ is an \emph{achievable secrecy rate} if for all $\delta>0$, there is an $n(\delta)\in\N$ and a sequence of $(n,M_n)$-codes $\{\sC_n\}_{n\in\N}$ with maximum error probability $e_n$ such that for all $n\geq n(\delta)$,
\begin{gather*}
	\frac{1}{n}\log M_n \geq R_s - \delta, \\
	\sup_{s\in\sS} I(M;Z_s^n)\leq\epsilon_n,
\end{gather*}
and $e_n,\epsilon_n\rightarrow0$ as $n\rightarrow\infty$. The \emph{compound secrecy capacity} $C_c$ of the wiretap channel $\comp$ is given by the supremum of all achievable secrecy rates $R_s$.
\end{definition}

\subsection{Countably-Finite Uncertainty Sets}
\label{sec:dmc_finite}

The discrete memoryless wiretap channel with a countably-finite uncertainty set (i.e. finite-state channels) is studied in \cite{Liang09CompoundWiretapChannels,Bjelakovic13CompoundWiretap}. In particular, the following achievable secrecy rate was established in \cite[Theorem 2]{Bjelakovic13CompoundWiretap}.

\begin{theorem}[\cite{Bjelakovic13CompoundWiretap}]
\label{the:finite}
The compound secrecy capacity $C_c$ of the discrete memoryless wiretap channel $\mathfrak{W}$ is lower-bounded as follows:
\begin{equation}
	C_c \geq \max_{P_X}\big(\min_{s\in\sS}I(X;Y_s)-\max_{s\in\sS}I(X;Z_s)\big)
	\label{eq:dmc_finiterate}
\end{equation}
where the uncertainty set $\sS$ is countably-finite. Here, the random variables $Y_s$ and $Z_s$ denote the outputs of the corresponding channels $W_s$ and $V_s$ for $s\in\sS$. \end{theorem}

Furthermore, it has been shown in \cite[Theorem 2]{Bjelakovic13CompoundWiretap} that the secrecy rate given in \eqref{eq:dmc_finiterate} is achieved with maximum probability of error of the form
\begin{equation}
	e_n \leq |\sS|^{1/4}2^{-n\alpha}
	\label{eq:dmc_finiteerror}
\end{equation}
and the secrecy constraint behaving as
\begin{equation}
	\max_{s\in\sS}I(M;Z_s^n)\leq 2^{-n\beta}
	\label{eq:dmc_strong}
\end{equation}
for some $\alpha,\beta>0$. Thus, both criteria, i.e., reliability \eqref{eq:dmc_error} and secrecy \eqref{eq:dmc_secrecy}, decrease \emph{exponentially fast} for increasing block length $n$. In addition, both bounds do \emph{not} depend on the particular channel realization. These two properties will be indispensable for extending this result from countably-finite to arbitrary uncertainty sets.

\subsection{Arbitrary Uncertainty Sets}
\label{sec:dmc_arbitrary}

The result above applies to finite-state channels (i.e., countably-finite uncertainty sets)  and discrete alphabets. Here, we extend it to arbitrary uncertainty sets, which are not required to be finite or countable. Subsequently, it will be extended to continuous alphabets and compact uncertainty sets in Section \ref{sec:continuous}.

To accomplish this, we adapt the ideas from Blackwell, Breiman, and Thomasian \cite{Blackwell59Compound} and approximate arbitrary compound wiretap channels by suitably chosen finite-state channels.

\begin{lemma}
\label{lem:comp1}
Let $\comp=\{(W_s,V_s):s\in\sS\}$ be a discrete memoryless wiretap channel with arbitrary uncertainty set $\sS$. For every integer $L\geq2|\sY|^2|\sZ|^2$, there is a compound wiretap channel $\comp_L=\{(\overline{W}_s,\overline{V}_s):s\in\sS_L\}$ with a countably-finite uncertainty set $\sS_L$,  $|\sS_L|\leq(L+1)^{|\sX||\sY||\sZ|}$, such that any $(W_s,V_s)\in\comp$ is closely approximated by some  $(\overline{W}_s,\overline{V}_s)\in\comp_L$ so that
\begin{itemize}
	\item[(a)] for all $x\in\sX$, $y\in\sY$, $z\in\sZ$,
\begin{subequations}
\label{eq:comp1_prob}
\begin{align}
|W_s(y|x)-\overline{W}_s(y|x)|&\leq |\sY||\sZ|/L, \label{eq:comp1_prob1}\\
|V_s(z|x)-\overline{V}_s(z|x)|&\leq |\sY||\sZ|/L, \label{eq:comp1_prob2}\\
W_s(y|x)&\leq2^{\frac{2|\sY|^2|\sZ|^2}{L}}\overline{W}_s(y|x), \\
V_s(z|x)&\leq2^{\frac{2|\sY|^2|\sZ|^2}{L}}\overline{V}_s(z|x).
\end{align}
\end{subequations}
	\item[(b)] For any input distribution $P_X\in\sP(\sX)$,
\begin{subequations}
\label{eq:comp1_mutual}
\begin{align}
|I(X;Y_s)-I(X;\overline{Y}_s)| &\leq 2(|\sY||\sZ|)^{3/2}/L^{1/2}, \\
|I(X;Z_s)-I(X;\overline{Z}_s)| &\leq 2(|\sY||\sZ|)^{3/2}/L^{1/2}. \label{eq:com1_mutualz}
\end{align}
\end{subequations}
\end{itemize}
\end{lemma}
\begin{IEEEproof}
The proof can be found in Appendix \ref{app:lem.1}.
\end{IEEEproof}
\vspace*{0.5\baselineskip}

This shows that we can approximate an arbitrary compound wiretap channel $\comp$ by a finite-state one $\comp_L$ so that any channel in $\comp$ is close in several senses to one of the new constructed channels in $\comp_L$ (they can be made arbitrary close by increasing the number of quantization levels $L$, which we exploit below). The next lemma shows that if there is a \emph{``good''} code for a wiretap channel, then the same code can be used for all wiretap channels in its neighborhood.

\begin{lemma}
\label{lem:comp2}
Let $(W_s,V_s)$ and $(\overline{W}_s,\overline{V}_s)$ be two wiretap channels and $L>0$ such that Lemma \ref{lem:comp1} holds. Then any $(n,M_n)$-code for $(\overline{W}_s,\overline{V}_s)$ is also an $(n,M_n)$-code for $(W_s,V_s)$ with
\begin{equation}
	e_n\leq2^{2n|\sY|^2|\sZ|^2/L}\bar{e}_n
	\label{eq:comp2_error}
\end{equation}
and
\begin{align}
	&|I(M;Z_s^n)- I(M;\overline{Z}_s^n)|\nonumber \\
	&\qquad\leq 4 n |\sY||\sZ|^2\log|\sZ|/L + 4 n H_2(|\sY||\sZ|^2/L).
	\label{eq:comp2_secrecy}
\end{align}
with $H_2(\cdot)$ the binary entropy function. Here, $e_n$ and $\bar{e}_n$ denote the maximum probabilities of error for the channels $W_s$ and $\overline{W}_s$ respectively, cf. \eqref{eq:dmc_error}.
\end{lemma}
\begin{IEEEproof}
The proof can be found in Appendix \ref{app:lem.2}.
\end{IEEEproof}

\begin{remark}
\label{rem:continuity}
The tight bound in \eqref{eq:comp2_secrecy} is established based on a recent result on the continuity of the secrecy capacity of compound wiretap channels \cite{BocheSchaeferPoorXXContinuity}, which in turn was established using a technique developed for quantum channels in \cite{Leung09ContinuityQuantumCapacities}. Following instead the classical approach of \cite{Blackwell59Compound} by applying \eqref{eq:com1_mutualz} leads to a loose bound
\begin{equation}
	|I(M;Z_s^n)- I(M;\overline{Z}_s^n)|\leq 2(|\sY||\sZ|)^{3n/2}/L^{1/2}
\end{equation}
which increases exponentially fast in the block length $n$. This then prohibits the proof of Theorem \ref{the:arbitrary} (using the number $L$ of quantization levels that scales up exponentially in $n$ would not help since the bound in \eqref{eq:dmc_approxerror} would diverge). Thus, using the tight bound in \eqref{eq:comp2_secrecy} (which increases only linearly in $n$) is essential. This bound also reveals the scaling of the number of quantization levels $L$ with $n$ required to make the approximation error arbitrary small. If only weak secrecy is of interest, then the normalized difference is bounded by a constant independent of $n$, which can be made as small as desired by using sufficiently large but fixed $L$, while strong secrecy requires $L$ to scale faster than $n \log n$ for the approximation error to become arbitrary small.
\end{remark}

The two lemmas above allow one to extend the finite-state result in Theorem \ref{the:finite} to arbitrary uncertainty sets. To proceed further, we need the following definitions that establish an ordering of compound wiretap channels.

\begin{definition}
\label{def:less noisy}
A compound DMC $V_{s_2}$ is said to be noisier than a compound DMC $W_{s_1}$ if
\begin{equation}
I(U;Y_{s_1}) \ge I(U;Z_{s_2})
\end{equation}
for any aggregate channel state $s=(s_1,s_2) \in \sS$, any random variable $U$ and any DMC $U \rightarrow X$ such that $U \rightarrow X \rightarrow (Y_{s_1},Z_{s_2})$ is a Markov chain.
\end{definition}

\begin{definition}
\label{def:degraded}
Compound DMC $V_{s_2}$ is said to be (physically) degraded with respect to compound DMC $W_{s_1}$ if $X \rightarrow Y_{s_1} \rightarrow Z_{s_2}$ is a Markov chain for any channel state $s=(s_1,s_2) \in \sS$ and any input $X$.
\end{definition}

These definitions are an extension of the corresponding definition for non-compound (single-state) channels, see e.g. \cite{CsiszarKoerner78BroadcastChannelsConfidentialMessages,Bloch11InformationTheoreticSecrecy}. Similarly to the single-state channels, it can be shown that ``degraded'' implies ``noisier'', but the converse is not true, i.e. the latter requirement is weaker than the former (so that there are channels that are ``noisier'' but not ``degraded''). An equivalent to the less noisy requirement, which is somewhat easier to verify, can be established in the same way as for the single-state channels (see \cite{vanDijk97SpecialClassBCC}).

\begin{proposition}
The compound DMC $V_{s_2}$ is noisier than the compound DMC $W_{s_1}$ if and only if $I(X;Y_{s_1})-I(X;Z_{s_2})$ is concave in the input distribution $P_X$ for any channel state $s=(s_1,s_2) \in \sS$.
\end{proposition}

The compound secrecy capacity of the discrete memoryless wiretap channel $\mathfrak{W}$ can now be characterized as follows.

\begin{theorem}
\label{the:arbitrary}
The compound secrecy capacity $C_c$ of the discrete memoryless wiretap channel $\mathfrak{W}$ is bounded as follows:
\begin{equation}
C_c \ge \sup_{P_X}\big(\inf_{s_1\in\sS_1}I(X;Y_{s_1}) - \sup_{s_2\in\sS_2} I(X;Z_{s_2})\big)
	\label{eq:dmc_rate}
\end{equation}
for any uncertainty set $\sS$ (not necessarily finite or countable), and the equality is attained if $V_{s_2}$ is noisier than  $W_{s_1}$.
\end{theorem}
\begin{IEEEproof}
The proof of the lower bound is based on Lemmas \ref{lem:comp1} and~\ref{lem:comp2}. We approximate the arbitrary compound wiretap channel $\mathfrak{W}$ by a finite-state one $\comp_L$ with the number of quantization levels  $L=L(n)$, which is selected in such a way that:
\begin{enumerate}
\item it satisfies the condition of Lemma 1,
\item the secrecy rate supported by the approximated channel approaches that of the original channel arbitrary closely (so that $L(n) \rightarrow \infty$ as $n \rightarrow\infty$),
\item maximum error probability approaches 0 as $n \rightarrow \infty$ (so that $L(n) > 2|\sY|^2|\sZ|^2/\alpha$ but  $L(n)$ has to increase slower than exponentially),
\item secrecy criterion approaches 0 as $n\rightarrow\infty$ (so that $L(n)$ has to increase faster than $n \log n$).
\end{enumerate}
Note that criterion 4) dictates the fastest increase of $L(n)$ and using the classical approach of \cite{Blackwell59Compound} would not satisfy it. The following analysis shows that $ L = a\cdot n^2$ is a proper choice for the number of quantization levels, where
\begin{equation}
a > 2|\sY|^2|\sZ|^2 \max\{1,1/\alpha\},
\label{eq:dmc_a}
\end{equation}
and $\alpha$ is as in \eqref{eq:dmc_finiteerror}.

For each $(W_s,V_s)\in\comp$, we select a sufficiently good approximation $(\overline{W}_s,\overline{V}_s)$ according to Lemma \ref{lem:comp1}. The corresponding finite-state compound channel is denoted by $\comp_L$ and the countably-finite uncertainty set by $\sS_L$, where $|\sS_L|\leq(L+1)^{|\sX||\sY||\sZ|}$.

Next, we check the reliability part. Fix input distribution $P_X$ and set the secrecy rate
\begin{equation}
R_s = \min_{s\in\sS} I(X;\overline{Y}_s)-\max_{s\in\sS}I(X;\overline{Z}_s)-\epsilon
\end{equation}
for some $\epsilon>0$. From Theorem \ref{the:finite}, there exists an $(n,M_n)$-code for $\mathfrak{W}_L$ with probability of error
\begin{align}\notag
	\bar{e}_n &\leq |\sS_L|^{1/4}2^{-n\alpha} \\
		&\leq (L+1)^{(|\sX||\sY||\sZ|)/4}2^{-n\alpha} \rightarrow 0 \ \mbox{as} \ n \rightarrow\infty,
\end{align}
where the steps follow from \eqref{eq:dmc_finiteerror}, $|\sS_L|\leq(L+1)^{|\sX||\sY||\sZ|}$, cf. Lemma \ref{lem:comp1}, and $L=a\cdot n^2$. Furthermore, from Lemma~\ref{lem:comp1}, for each $W_s \in \mathfrak{W}$ there is an appropriate $\overline{W}_s \in \mathfrak{W}_L$ such that $W_s(y|x)\leq2^{2|\sY|^2|\sZ|^2/L}\overline{W}_s(y|x)$ for all $x,y$. Thus, Lemma \ref{lem:comp2} implies that the code for $\mathfrak{W}_L$ is also a code for $\mathfrak{W}$ with probability of error
\begin{align}
	e_n &\leq 2^{n\frac{2|\sY|^2|\sZ|^2}{L}}\bar{e}_n \nonumber \\
		&\leq |\sS_L|^{1/4}2^{-n(\alpha-\frac{2|\sY|^2|\sZ|^2}{L})} \nonumber \\
		&\leq (L+1)^{(|\sX||\sY||\sZ|)/4}2^{-n(\alpha-\frac{2|\sY|^2|\sZ|^2}{L})}.
		\label{eq:dmc_approxerror}
\end{align}
Since $L=an^2$, we have $e_n\rightarrow0$ as $n\rightarrow\infty$. This means the code constructed for the approximated channel $\comp_L$ also satisfies the reliability criteria for the original channel $\comp$. Thus, it remains to show that the rate of this code is arbitrarily close to the desired rate and the strong secrecy condition is satisfied. From Lemma \ref{lem:comp1}(b), one obtains, for any input distribution $P_X$,
\begin{align}
	 &\Big|\inf_{s\in\sS}I(X;Y_s)-\sup_{s\in\sS}I(X;Z_s) \nonumber \\ &\qquad\qquad-\big(\min_{s\in\sS}I(X;\overline{Y}_s)-\max_{s\in\sS}I(X;\overline{Z}_s)\big)\Big| \nonumber \\
	 &\qquad\leq 4(|\sY||\sZ|)^{3/2}/L^{1/2}.
\label{eq:dmc_approxrate}
\end{align}
Thus, the difference between the rate achieved by the code for the approximated finite-state channel $\comp_L$ and the desired rate for the original channel $\comp$ is arbitrarily small, since $L\rightarrow\infty$ as $n\rightarrow\infty$.

It remains to check that the secrecy constraint is also satisfied. The code above for the approximated finite-state channel $\mathfrak{W}_L$ has $\max_{s\in\sS_L}I(M;\overline{Z}_s^n)\leq2^{-n\beta}$, cf. Theorem \ref{the:finite} and \eqref{eq:dmc_strong}, so that evoking Lemma \ref{lem:comp2}, one obtains
\begin{align}
	\sup_{s\in\sS}I(M;Z_s^n) &\leq \max_{s\in\sS}I(M;\overline{Z}_s^n) + 4 n |\sY||\sZ|^2\log|\sZ|/L \nonumber\\
	&\qquad+ 4 n H_2(|\sY||\sZ|^2/L) \nonumber \\
	&\leq 2^{-n\beta} + 4|\sY||\sZ|^2\log|\sZ|/(a n) \nonumber\\
	&\qquad+ 4 n H_2(|\sY||\sZ|^2/(a n^2)) \nonumber \\
&\rightarrow 0 \ \mbox{as} \ n \rightarrow\infty
	\label{eq:dmc_approxsecrecy}
\end{align}
where we have used $n H_2(\alpha/n^2) \rightarrow 0$ as $n\rightarrow \infty$ for any $\alpha>0$. Thus, also the information leaked to the eavesdropper is arbitrarily small.

To establish the equality part under the noisier condition, observe that, by extending the proof of the converse in Theorem 3 of \cite{CsiszarKoerner78BroadcastChannelsConfidentialMessages} to the compound setting and requiring the encoder to be independent of the actual channel states, it can be shown that any achievable secrecy rate is bounded as follows
\begin{align}
\label{eq.T2.ub1}
R_s \le I(X;Y_{s_1}) - I(X;Z_{s_2})
\end{align}
for any channel state $(s_1,s_2)$, where the input $X$ is induced by the encoder, so that
\begin{align}
R_s \le \inf_s I(X;Y_{s_1}) - I(X;Z_{s_2})
\end{align}
from which it follows that
\begin{align}
\label{eq.T2.ub3}
C_c \le \sup_{P_X} \inf_s I(X;Y_{s_1}) - I(X;Z_{s_2})
\end{align}
and thus establishes the equality.
\end{IEEEproof}

\begin{remark}
\label{rem:approximation}
The proof of Theorem \ref{the:arbitrary} reveals that the required scaling of $L(n)$ depends on the secrecy criterion adopted, cf. in particular \eqref{eq:dmc_approxsecrecy}: this requires $L(n)$ to scale faster than $n \log n$ and motivates the convenient choice of $L(n) = a \cdot n^2$ for some $a$ satisfying \eqref{eq:dmc_a} (in fact, using $n^{1+\delta}$ with any $\delta >0$ would work as well).  Requiring weak secrecy instead allows for the quantization number $L(n)$ to increase arbitrarily slowly in $n$ (e.g. as $\log n$ or $\log\log n$).
\end{remark}

\begin{remark}
\label{rem:exponential}
It should be emphasized that two properties of the probability of error \eqref{eq:dmc_finiteerror} and the secrecy \eqref{eq:dmc_strong} are indispensable to extend the result from finite uncertainty sets to arbitrary uncertainty sets: its exponentially-fast decreasing behavior and its independence of the actual channel realization. Thus, such bounds have to be established carefully for the finite case since otherwise an extension to the arbitrary case is not possible. Moreover, the approximation must be done carefully enough (e.g. as in \eqref{eq:comp2_secrecy} with $L(n)=a\cdot n^2$) to ensure that both the secrecy and reliability criteria are still valid after the approximation.
\end{remark}

\begin{remark}
Since each degraded channel is also ``noisier'', the equality in Theorem \ref{the:arbitrary} also holds for degraded channels.
\end{remark}

To proceed further, we need the following definitions.

\begin{definition}
\label{def:less capable}
Compound DMC $V_{s_2}$ is said to be less capable than compound DMC $W_{s_1}$  if for every $P_X$ and any channel state $(s_1,s_2) \in \sS$
\begin{equation}
I(X;Y_{s_1}) \ge I(X;Z_{s_2}).
\end{equation}
\end{definition}

This definition extends the corresponding definition in \cite{Bloch11InformationTheoreticSecrecy} to the compound channel setting. Following the same line of analysis as for single-state channels, it can be shown that the less capable requirement is strictly weaker than the noisier one (i.e. each ``noisier'' channel is also ``less capable'' but the converse is not true), and hence strictly weaker than the degraded one.

\begin{definition}
\label{def:saddle point}
A compound wiretap channel is said to have a saddle-point if
\begin{equation}
\label{eq:saddle point}
\begin{split}
&\sup_{P_X}\inf_{s\in\sS}\big(I(X;Y_{s_1}) - I(X;Z_{s_2})\big) \\
	&\qquad\qquad= \inf_{s\in\sS} \sup_{P_X}\big(I(X;Y_{s_1}) - I(X;Z_{s_2})\big)
\end{split}
\end{equation}
where $s=(s_1,s_2)$ is the aggregate channel state.
\end{definition}

Note that this definition does not impose any operational meaning on the quantities involved. The following corollary provides such operational meaning.

\begin{corollary}
\label{cor:saddle}
If the compound wiretap channel $\mathfrak{W}$ has a saddle-point and satisfies the less capable condition, then the compound secrecy capacity $C_c$ is the same as the worst-case channel capacity $C_w$,
\begin{equation}
\label{eq:C_c cor}
\begin{split}
	C_c &= \sup_{P_X}\big(\inf_{s_1\in\sS_1}I(X;Y_{s_1}) - \sup_{s_2\in\sS_2} I(X;Z_{s_2})\big) \\
	&= \inf_{s\in\sS} \sup_{P_X}\big(I(X;Y_{s_1}) - I(X;Z_{s_2})\big) = C_w.
\end{split}
\end{equation}
In particular, the channel has a saddle-point if
\begin{enumerate}
\item  $\sS_1, \sS_2$ are compact and convex, and
\item $I(X;Y_{s_1}) - I(X;Z_{s_2})$ is lower semi-continuous and quasi-convex  in $s$, and upper semi-continuous and quasi-concave in $P_X$.
\end{enumerate}
\end{corollary}
\begin{IEEEproof}
Since the legitimate and eavesdropper channel states are independent of each other, it follows that
\begin{align}
&\inf_{s_1\in\sS_1}I(X;Y_{s_1}) - \sup_{s_2\in\sS_2} I(X;Z_{s_2}) \nonumber\\
	&\qquad\qquad=
\inf_{s\in\sS}\big(I(X;Y_{s_1}) - I(X;Z_{s_2})\big)
\end{align}
so that the following chain inequality holds
\begin{align} \notag
C_w &= \inf_{s\in\sS}\sup_{P_X} \big(I(X;Y_{s_1}) - I(X;Z_{s_2})\big) \\ \notag
&= \sup_{P_X} \inf_{s\in\sS} \big(I(X;Y_{s_1}) - I(X;Z_{s_2})\big) \\ \notag
&\le C_c \\
&\le C_w
\end{align}
where first equality holds since, from \cite[Corollary 3.5]{Bloch11InformationTheoreticSecrecy},
\begin{align}
\sup_{P_X} \big(I(X;Y_{s_1}) - I(X;Z_{s_2})\big)
\end{align}
is the secrecy capacity under channel state $(s_1,s_2)$ and the less capable condition, so that taking $\inf_s$ gives the worst-case capacity; the first inequality is due to Theorem \ref{the:arbitrary} and the last inequality is due to the fact that compound capacity cannot exceed the worst-case one (since the compound code has also to work on the worst-case channel). This proves $C_c=C_w$. The last statement follows from von Newmann mini-max theorem and its subsequent generalizations, see e.g. \cite[Theorem 9.D]{Zeidler86NonlinearFunctionalAnalysisI}.
\end{IEEEproof}

\begin{remark}
The importance of this result is due to the fact that a code designed for the worst-case channel also works on the whole class of channels , i.e. is robust (which is not true in general).
\end{remark}

\begin{remark}
The requirement of semi-continuity can be dropped in the case of countably-finite alphabets (since the mutual information is known to be continuous in such settings), but it is essential for countably-infinite or continuous alphabets.
\end{remark}

\begin{remark}
Since ``noisier'' implies ``less capable'', Corollary \ref{cor:saddle} also holds for ``noisier'' and degraded (physically or stochastically) channels.
\end{remark}

Thus, Theorem \ref{the:arbitrary} extends Theorem \ref{the:finite} to arbitrary uncertainty sets. The next step is to extend this result to continuous alphabets.

\section{Continuous Alphabets}
\label{sec:continuous}

To establish an achievable secrecy rate for the compound Gaussian MIMO wiretap channel, we have to deal with continuous input and output alphabets as well as probability density functions. Therefore, we extend the previous result in Theorem \ref{the:arbitrary} to continuous alphabets.

Let us consider the general case of input and output alphabets $\sX$, $\sY$, and $\sZ$ which are standard \cite{Gray09Probability}. Such alphabets include practically relevant cases such as continuous alphabets in Euclidean spaces or finite alphabets (see \cite{Gray09Probability} and \cite{Gray11Entropy} for an extensive discussion of this; the requirement of random variables to be defined over a standard space ensures that conditional probability measures are well-defined). Accordingly, we assume that the corresponding random variables can be described by probability density functions and that all mutual information terms are calculated according to continuous alphabets and are finite.

Usually, results are extended from discrete memoryless channels to continuous channels by using the \emph{discretization procedure} or \emph{partitioning method} as outlined for example in \cite{ElGamal11NetworkInformationTheory}; see \cite{Gray11Entropy} or \cite{Gallager68InformationTheoryReliableCommunication} respectively for a more detailed treatment. Such an approach invokes quantization arguments, where for any input distribution $p_X$ for the continuous channel, the input and output alphabets are partitioned making the results for finite alphabets applicable. Letting the corresponding quantizer be sufficiently fine, the actual mutual information terms of the partitioned alphabets can be made arbitrarily close to the continuous one.

Applying this approach to compound channels has to be done carefully. We have to ensure that the sequence of successively finer quantizers partitions the input and output alphabets in such a way that the mutual information terms between the quantized alphabets approaches the desired terms for continuous alphabets for all possible channel realizations simultaneously. Thus, the invoked quantizers must not depend on a particular channel realization. This issue is discussed in detail in \cite{Mitran06CompoundSideInformation} which studies the compound channel with side information. The following result is a slight extension of the corresponding result in \cite{Mitran06CompoundSideInformation} to the wiretap channel setting.

\begin{lemma}
\label{lem:cont1}
For the compound wiretap channel $\comp$ with standard input and output alphabets, there exists a sequence of successively finer quantizers $\{q_{X,k},q_{Y,k},q_{Z,k}\}_{k\in\N}$ for the input and outputs such that for any channel realization $s\in\sS$
\begin{align*}
	I(X;Y_s) &= \lim_{k\rightarrow\infty}I\big(q_{X,k}(X);q_{Y,k}(Y_s)\big) \\
	I(X;Z_s) &= \lim_{k\rightarrow\infty}I\big(q_{X,k}(X);q_{Z,k}(Z_s)\big).
\end{align*}
This means there exist universal sequences of quantizers which work for all channel realizations $s\in\sS$ simultaneously if the input and output alphabets are standard.
\end{lemma}
\begin{IEEEproof}
See \cite[Lemma 3]{Mitran06CompoundSideInformation} and also \cite{Gray11Entropy} for further details.
\end{IEEEproof}
\vspace*{0.5\baselineskip}

The second technicality is that one has to ensure that such sequences of functions converge uniformly on a compact set when they converge pointwise (this is needed since the transmitter does not know the channel state).

\begin{lemma}
\label{lem:cont2}
Let $W_s, V_s$ be continuously parametrized by $s \in \sS$, where $\sS$ is a compact set. Then, for all channel realizations $s\in\sS$ and for each input distribution $p_X$, there exists a sequence of successively finer universal quantizers $\{q_{X,k},q_{Y,k},q_{Z,k}\}_{k\in\N}$ such that for each $\epsilon>0$, there is an $n(\epsilon)\in\N$ such that for every $k\geq n(\epsilon)$,
\begin{align*}
	&I\big(q_{X,k}(X);q_{Y,k}(Y_s)\big)-	 I\big(q_{X,k}(X);q_{Z,k}(Z_s)\big) \\
	&\qquad\qquad\qquad\qquad\geq \inf_{s\in\sS}I(X;Y_s)-\sup_{s\in\sS}I(X;Z_s)-\epsilon.
\end{align*}
\end{lemma}
\begin{IEEEproof}
The proof follows by applying \cite[Lemmas 4 and 5]{Mitran06CompoundSideInformation} to both terms $I(q_{X,k}(X);q_{Y,k}(Y_s))$ and $I(q_{X,k}(X);q_{Z,k}(Z_s))$.
\end{IEEEproof}
\vspace*{0.5\baselineskip}

Having these technicalities in mind, we are now in the position to establish the desired result for continuous alphabets.

\begin{theorem}
\label{the:continuous}
The compound secrecy capacity $C_c$ of the wiretap channel $\mathfrak{W}$ continuous in $s$ and with standard (possibly continuous) input and output alphabets is bounded as follows:
\begin{equation}
C_c \ge \sup_{p_X}\big(\inf_{s_1\in\sS_1}I(X;Y_{s_1}) - \sup_{s_2\in\sS_2} I(X;Z_{s_2})\big)
	\label{eq:dmc_continuousrate}
\end{equation}
for any compact uncertainty set $\sS$, and the equality is attained if $V_{s_2}$ is noisier than $W_{s_1}$.
\end{theorem}
\begin{IEEEproof}
To prove the lower bound, we follow the discretization procedure and use sequences of successively finer quantizers $\{q_{X,k},q_{Y,k},q_{Z,k}\}_{k\in\N}$ according to Lemmas~\ref{lem:cont1} and \ref{lem:cont2} which partition the continuous input and output alphabets in such a way that we end up with mutually disjoint events which cover the entire spaces. Then, all mutual information terms are calculated according to these partitions.

For each choice of quantizers $q_{X,k},q_{Y,k},q_{Z,k}$, the whole encoding and decoding procedure as used in the proofs of Theorems \ref{the:finite} and \ref{the:arbitrary}, cf. also \cite{Bjelakovic13CompoundWiretap}, is done according to this partition. Then, the analysis of probability of error and the analysis of the secrecy criterion for finite alphabets ensures that any rate $R_s$ satisfying
\begin{align*}
	&R_s < \sup_{P_X}\big(\inf_{s\in\sS}I\big(q_{X,k}(X);q_{Y,k}(Y_s)\big) \\
	&\qquad\qquad\qquad\qquad\qquad- \sup_{s\in\sS}I\big(q_{X,k}(X);q_{Z,k}(Z_s)\big)
\end{align*}
is achievable with strong secrecy for the compound wiretap channel.

From Lemmas \ref{lem:cont1} and \ref{lem:cont2}, for standard alphabets $\sX$, $\sY$, $\sZ$, and any $\epsilon>0$, one can find sequences of successively finer quantizers $\{q_{X,k},q_{Y,k},q_{Z,k}\}_{k\in\N}$ such that for sufficiently large $n$ and any $k \ge n$,
\begin{equation}
\begin{split}
	&\inf_{s\in\sS}I\big(q_{X,k}(X);q_{Y,k}(Y_s)\big)-	 \sup_{s\in\sS}I\big(q_{X,k}(X);q_{Z,k}(Z_s)\big) \\
	&\qquad\qquad\qquad\geq \inf_{s\in\sS}I(X;Y_s)-\sup_{s\in\sS}I(X;Z_s)-\epsilon
	\label{eq:dmc_partition}
\end{split}
\end{equation}
so that any rate
\begin{equation*}
R_s \le \sup_{p_X}(\inf_{s\in\sS}I(X;Y_s) - \sup_{s\in\sS}I(X;Z_s))-\epsilon
\end{equation*}
is achievable for standard (continuous) alphabets as well, from which \eqref{eq:dmc_continuousrate} follows. Note that as the uncertainty set is assumed to be compact and therewith bounded, all terms are well defined and finite for standard alphabets as well. The equality part is established as in Theorem 2 (using the upper bounds in \eqref{eq.T2.ub1}-\eqref{eq.T2.ub3}, which apply to continuous alphabets as well). This completes the proof.
\end{IEEEproof}
\vspace*{0.5\baselineskip}

Using this theorem, Corollary 1 can be extended to continuous alphabets in a natural way.

\section{Gaussian MIMO Channels}
\label{sec:mimo}

We are now in the position to specialize the result in Theorem~\ref{the:continuous} to Gaussian MIMO channels. To this end, let $N_T$ be the number of transmit antennas at the transmitter and $N_{1(2)}$ be the numbers of receive antennas at the legitimate receiver (eavesdropper). The input-output relations for the Gaussian MIMO wiretap channel are given by
\begin{equation}
\label{eq.M1}
\by_1 =\bH_1\bx + \bxi_1 , \quad \by_2 = \bH_2 \bx + \bxi_2
\end{equation}
where $\bx=[x_1 ,x_2 ,...,x_{N_T} ]^T\in\C^{N_T\times1}$ is the transmitted signal, $\by_{1(2)}\in\C^{N_{1(2)}\times1}$ is the signal at the legitimate receiver (eavesdropper), $\bxi_{1(2)}\in\C^{N_{1(2)}\times1}$ is the circularly-symmetric additive white Gaussian noise at the receiver (eavesdropper) (normalized to unit variance in each dimension), and $\bH_{1(2)}\in\C^{N_{1(2)}\times N_T}$ is the matrix of the complex channel gains between each transmit and each receive (eavesdropper) antenna. The channels ${\rm {\bf H}}_{1(2)} $ are assumed to be fixed (constant) during the whole transmission of block length $n$. We assume an average transmit power constraint $\tr\bR\leq P_T$ where $P_T$ is the total transmit power and $\bR=\E\{\bx\bx^+\}$ is the transmit covariance matrix.

For this channel, the secrecy capacity subject to the total average transmit power constraint is \cite{Liu09MIMOWiretapSecrecy,KhistiWornell10MIMOWiretap1,KhistiWornell10MIMOWiretap2,Oggier11MIMOWiretap}
\begin{equation}
\label{eq.M2}
C_s =\mathop {\max }\limits_{\bR} \ln \frac{\left| \bI + \bW_1 \bR \right|}{\left| \bI+\bW_2 \bR\right|}
\end{equation}
where $\bW_i =\bH_i^+ \bH_i$, $i=1,2$, and $\max$ is subject to the constraints $\bR\geq\vec{0}$ and $\tr\bR\leq P_T$.

It is well-known that the problem in \eqref{eq.M2} is not convex in general and explicit solutions for the optimal transmit covariance are not known for the general case, but only for some special cases (e.g. low-SNR, MISO channels, or for the full-rank case) \cite{Liu09MIMOWiretapSecrecy,KhistiWornell10MIMOWiretap1,KhistiWornell10MIMOWiretap2,Oggier11MIMOWiretap,Loyka12OptimalSignalingMIMOWiretap}.

Let us now consider a compound Gaussian MIMO wiretap channel where the exact channel realizations $\bH_1$ and $\bH_2$ are unknown. It is only known to the legitimate user that they belong to the compact set $\vec{\sS}$. Again, we make the safest assumption from the secrecy perspective and assume that the eavesdropper knows both $\bH_1$ and $\bH_2$, cf. also Remark \ref{rem:csi}. Then, evaluating Theorem \ref{the:continuous} for this particular choice of compound Gaussian MIMO channel yields the following achievable secrecy rate.

\begin{corollary}
\label{cor:mimo}
The (strong) compound secrecy capacity $C_c$ of the compound Gaussian MIMO channel in \eqref{eq.M1} is lower-bounded as follows:
\begin{equation*}
	C_c \geq \max_{\bR}\min_{\bW_1,\bW_2}\ln \frac{\left| \bI + \bW_1 \bR \right|}{\left| \bI+\bW_2 \bR\right|}
\end{equation*}
where $\max$ and $\min$ are subject to  $\bR,\bW_1,\bW_2\geq\vec{0}$, $\tr\bR\leq P_T$, and $\bW_1$, $\bW_2$ belong to a compact set $\vec{\sS}$.
\end{corollary}

A similar result was given earlier in \cite[Lemma 1]{Liang09CompoundWiretapChannels} under the weak secrecy constraint and finite-state channels (countably-finite uncertainty sets). Corollary \ref{cor:mimo} extends it to strong secrecy and arbitrary (compact) uncertainty sets.

\section{Eavesdropper Channel Uncertainty}
\label{sec:uncert}

Let us now consider a particular compound channel where $\bH_1$ is given (known to the transmitter) and $\bH_2$ can be any (unknown) subject to the spectral norm constraint
\begin{align}
\begin{split}
\label{eq.M3}
	\sSt =&{} \big\{\bH_2:|\bH_2|_2 = \max_{|\bx|=1} |\bH_2\bx| \le \sqrt{\epsilon}\big\} \\
		=&{}\big\{\bW_2:|\bW_2|_2 = \lambda_1(\bW_2) \le \epsilon\big\}
\end{split}
\end{align}
where $|\bx| = \sqrt{\bx^+\bx}$ is the Euclidean norm of $\bx$,
$|\bH|_2=\sigma_1(\bH)$ is the spectral norm of $\bH$, i.e. its  largest singular value $\sigma_1(\bH)$; $\lambda_1(\bW_2)$ is the largest eigenvalue of $\bW_2$. Thus, the set $\vec{\sS}_2$ includes \emph{all} $\bW_2$ that are less than or equal to $\epsilon\bI$.

Note that $|\bH\bx|$ represents the channel (voltage) gain in transmit direction $\bx$ so that $|\bH|_2$ is the largest channel gain. $|\bW|_2$ represents the largest channel power gain. The importance of the spectral norm in the context of regular MIMO channels has been discussed in \cite{Loyka12CompoundMIMO}. Essentially the same motivation applies to the secure MIMO channel here. In particular, the set in \eqref{eq.M3} limits the maximum gain of the eavesdropper channel without putting any constraint on its eigenvectors. This represents the physical scenario where the eavesdropper cannot approach the transmitter beyond a certain minimum (protection) distance (so that the channel gain is bounded due to propagation path loss) being unconstrained  otherwise.

To establish the secrecy capacity of this compound channel (not necessarily degraded) in Theorem \ref{thm.M1}, we establish first a number of intermediate results in Propositions \ref{prop.M1} and \ref{prop.M2}.

\subsection{Worst-Case Secrecy Capacity and Saddle-Point Property}
\label{sec:uncert_worstcase}

The following proposition gives the capacity of the worst-case channel in this set. For this purpose we define
\begin{align}
C(\bR,\bW_2) = \ln \frac{\left| \bI + \bW_1 \bR \right|}{\left| \bI+\bW_2 \bR\right|}
\end{align}
which depends on the transmit covariance matrix $\bR$ and the eavesdropper channel $\bW_2=\bH_2^+\bH_2$ unknown to the transmitter. The channel to the legitimate receiver $\bW_1=\bH_1^+\bH_1$ is fixed and known to the transmitter.

\begin{proposition}
\label{prop.M1}
Consider the class of channels in \eqref{eq.M1} for a given (known) $\bW_1$ and any $\bW_2 \in \sSt$ (as in \eqref{eq.M3}). Then, the secrecy capacity $C_w$ of a worst-case channel is
\begin{align}
\label{eq.prop.M1}
C_w = \min_{\bW_2} \mathop {\max }\limits_{\bR} C(\bR,\bW_2) = C^\ast(\epsilon)
\end{align}
where $\max$ and $\min$ are over all admissible $\bR, \bW_2$:  $\bR, \bW_2 \ge {\bf 0}$, $\tr \bR \le P_T$, $\bW_2\in\sSt$, i.e. $|\bW_2|_2 \le \epsilon$, and
\begin{align}
C^\ast(\epsilon) = \mathop {\max }\limits_{\tr \bR \le P_T} C(\bR,\epsilon \bI)
\end{align}
is the secure capacity for the isotropic eavesdropper $\bW_{2w}= \epsilon \bI$, which is the worst-case eavesdropper in $\sSt$.
\end{proposition}
\begin{IEEEproof}
Observe that $|\bI + \bW \bR|$ is monotonically increasing in $\bW$, i.e.
\[
|\bI + \bW_1 \bR| \ge |\bI + \bW_2 \bR| \ \mbox{if} \ \bW_1 \ge \bW_2
\]
(see e.g. \cite{Zhang11MatrixTheory}), so that
\[
C(\bR, \bW_2) \ge C(\bR, \epsilon \bI)
\]
for any $\bR$, with equality if $\bW_2 = \epsilon \bI$. Taking $\min\max$ of both parts results in \eqref{eq.prop.M1}.
\end{IEEEproof}
\vspace*{0.5\baselineskip}

It follows from Proposition \ref{prop.M1} that the isotropic eavesdropper is the worst-case one under a bounded channel gain for \textit{any} $\bW_1$. This is also appealing from the channel feedback perspective: it is hardly possible to expect that the eavesdropper will share its channel with the transmitter to make eavesdropping harder, so only minimal information can be expected by the transmitter about the eavesdropper channel.

\begin{figure}
\centerline{\includegraphics[width=3in]{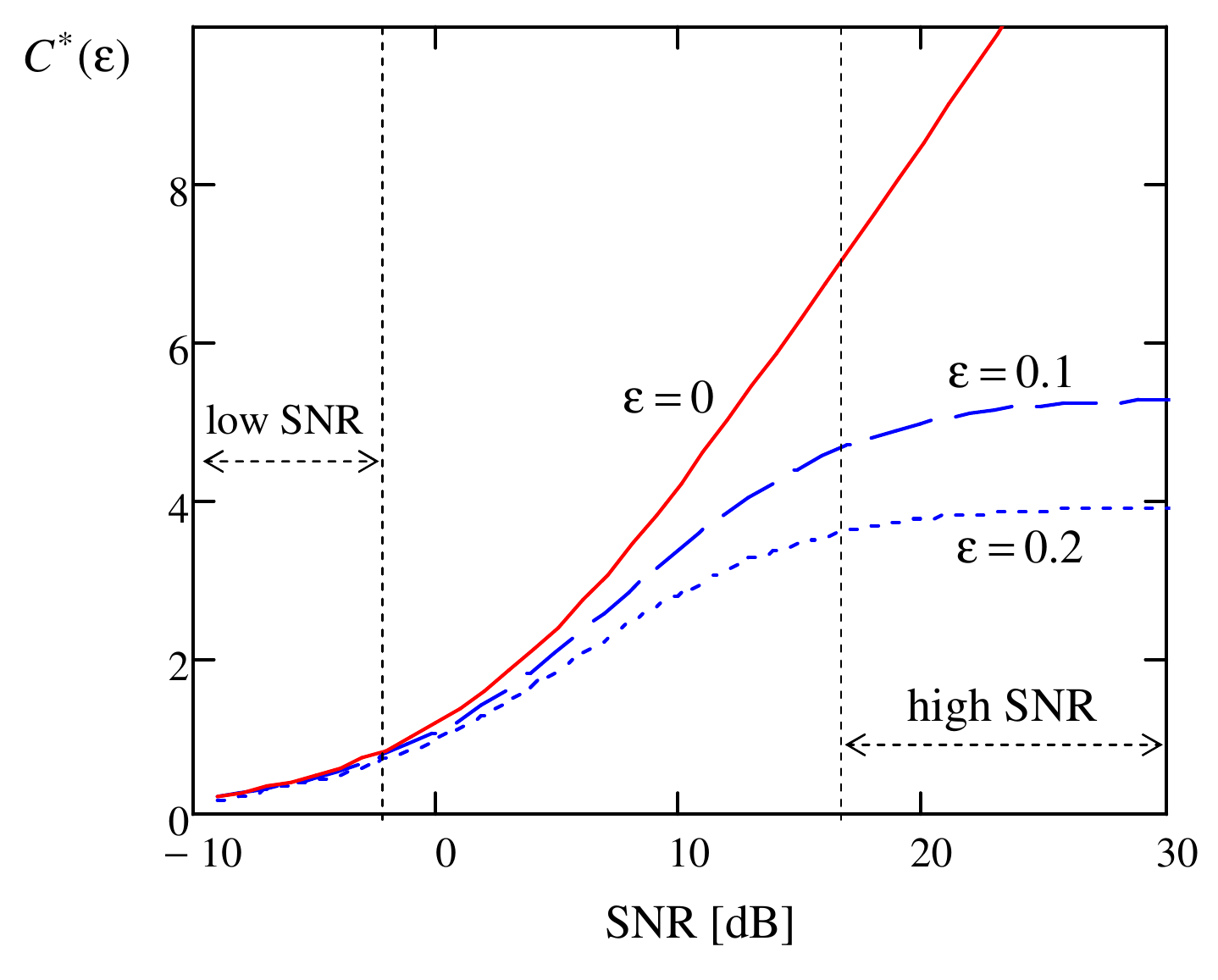}}
\caption{Secrecy capacity for the isotropic eavesdropper and the capacity of the regular MIMO channel (no eavesdropper, $\epsilon=0$) vs. the SNR ($=P_T$ since the noise variance is unity); $\lambda_1(\bW_1)=2, \ \lambda_2(\bW_1)=1$. Note the saturation effect at high SNR, where the capacity strongly depends on $\epsilon$ but not the SNR, and the negligible impact of the eavesdropper at low SNR.}
\label{fig:Fig_1}
\end{figure}

The secrecy capacity $C^\ast(\epsilon)$ under the isotropic eavesdropper has been studied in details in \cite{Loyka13FurtherResultsSecureMIMO}, including its high/low SNR approximations and capacity bounds for the general (non-isotropic) case. In particular, $C^\ast(\epsilon)$ is a decreasing, convex function of $\epsilon$. As Fig. \ref{fig:Fig_1} shows, the presence of eavesdropper results in capacity saturation at high SNR, where the eavesdropper's impact is much more pronounced.

The following proposition demonstrates the saddle-point property for the class of channels in \eqref{eq.M3} which will be important later to prove the converse result for the compound secrecy capacity.

\begin{proposition}
\label{prop.M2}
Consider the class of channels in \eqref{eq.M1} for a given (known) $\bW_1$ and any $\bW_2 \in \sSt$. The following saddle-point property holds:
\begin{align}
\mathop {\max }\limits_{\bR} \min_{\bW_2} C(\bR,\bW_2) = \min_{\bW_2} \mathop {\max }\limits_{\bR} C(\bR,\bW_2)
\end{align}
where $\max$ and $\min$ are over all admissible $\bR, \bW_2$.
\end{proposition}
\begin{IEEEproof}
For the max-min part, observe that $C(\bR,\bW_2) \ge C(\bR,\epsilon\bI)$ (which follows from the proof of Proposition \ref{prop.M1}), so by taking $\max\min$ of both parts, one obtains
\begin{align}
\mathop {\max }\limits_{\bR} \min_{\bW_2} C(\bR,\bW_2) \ge \mathop {\max }\limits_{\bR} C(\bR,\epsilon\bI).
\end{align}
On the other hand, by using $\bW_2 = \epsilon\bI$ instead of $\min$, one obtains
\begin{align}
\mathop {\max }\limits_{\bR} \min_{\bW_2} C(\bR,\bW_2) \le \mathop {\max }\limits_{\bR} C(\bR,\epsilon\bI)
\end{align}
so that
\begin{align} \notag
\mathop {\max }\limits_{\bR} \min_{\bW_2} C(\bR,\bW_2) &= \mathop {\max }\limits_{\bR} C(\bR,\epsilon\bI) \\
&= \min_{\bW_2} \mathop {\max }\limits_{\bR} C(\bR,\bW_2).
\end{align}
This proves the desired saddle-point property.
\end{IEEEproof}

\subsection{Compound Secrecy Capacity}
\label{sec:uncert_capacity}

The saddle-point property above is instrumental in establishing the secrecy capacity of the compound MIMO channel in \eqref{eq.M1} and \eqref{eq.M3} as the following theorem  shows.

\begin{theorem}
\label{thm.M1}
Consider the compound Gaussian MIMO wiretap channel in \eqref{eq.M1} with known $\bW_1$ and unknown $\bW_2$ belonging to the uncertainty set $\sSt$ in \eqref{eq.M3}. Its compound secrecy capacity $C_c$ is
\begin{align} \notag
\label{eq.M.T1.1}
C_c &= \mathop {\max }\limits_{\bR} \min_{\bW_2} C(\bR,\bW_2) \\
\notag
&= \min_{\bW_2} \mathop {\max }\limits_{\bR} C(\bR,\bW_2) \\
&= C^\ast(\epsilon)
\end{align}
where $\max$ and $\min$ are over all admissible $\bR, \bW_2$. The optimal signaling is Gaussian and on the eigenmodes of the legitimate channel,
\begin{align}
\label{eq.M.T1.2}
{\bf R}^* = {\bf U}_1 {\bf \Lambda}^*{\bf U}_1^+,
\end{align}
where the columns of unitary matrix ${\bf U}_1$ are the eigenvectors of ${\bf W}_1$, diagonal matrix ${\bf \Lambda} = \diag\{\lambda_i^*\}$ collects the eigenvalues of ${\bf R}^*$,
\begin{eqnarray}
\label{eq.M.T1.3}
\lambda_i^* = \frac{\epsilon + g_i}{2\epsilon g_i} z_i, \ z_i= \sqrt{1+ \frac{4\epsilon g_i}{(\epsilon + g_i)^2} \left( \frac{g_i - \epsilon}{\lambda} - 1 \right)_+} -1
\end{eqnarray}
and $\lambda > 0$ is found from the total power constraint $\sum_i \lambda_i^* = P_T$, $g_i=\lambda_i({\bf W}_1)$, $(x)_+ = \max\{x,0\}$. The secrecy capacity can be expressed as
\bal
\label{eq.M.T1.4}
C^*(\epsilon) = \sum_i \ln \frac{1 + g_i \lambda_i^*}{1 + \epsilon \lambda_i^*} = \sum_{i_+} \ln \frac{g_i}{\epsilon} + \sum_{i_+} \ln \frac{2\epsilon + (\epsilon+g_i)z_i}{2g_i + (\epsilon+g_i)z_i}
\eal
where the summation is over the set of active eigenmodes:
\bal
i_+=\{i: g_i > \lambda + \epsilon\}.
\eal
\end{theorem}
\begin{IEEEproof}
Note first that
\begin{align}
C_c \le \min_{\bW_2} \mathop {\max }\limits_{\bR} C(\bR,\bW_2),
\end{align}
i.e., the compound capacity cannot exceed the worst-case capacity in the class and the latter is achieved by Gaussian signaling. On the other hand, it follows from Corollary \ref{cor:mimo} that
\begin{align}
C_c &\ge \mathop {\max }\limits_{\bR} \min_{\bW_2} C(\bR,\bW_2) \\
&= \min_{\bW_2} \mathop {\max }\limits_{\bR} C(\bR,\bW_2)
\end{align}
where the equality is from Proposition \ref{prop.M2}. Combining the lower and upper bounds, \eqref{eq.M.T1.1} and optimality of Gaussian signaling for the compound channel follow. The optimal covariance in \eqref{eq.M.T1.2}-\eqref{eq.M.T1.3} and the capacity in \eqref{eq.M.T1.4} follow from Proposition 2 in \cite{Loyka13FurtherResultsSecureMIMO} since the worst-case eavesdropper is isotropic.
\end{IEEEproof}
\vspace*{0.5\baselineskip}

Note that this theorem does not require the compound channel to be degraded (as is the case for the known capacity results, where all eavesdropper channel states are required to be degraded with respect to all legitimate user channel states). It shows that the secrecy capacity of the worst-case channel is also the (compound) secrecy capacity of the class of channels (achievable by a single code on the whole class) so that Gaussian signaling is optimal, and the following saddle-point inequalities hold for any feasible $\bR$ and $\bW_2$,
\begin{align}
\label{eq.M.saddle point}
C(\bR,\epsilon\bI) \le C_c = C(\bR^*,\epsilon\bI) \le C(\bR^*,\bW_2)
\end{align}
where $(\bR^*, \epsilon\bI)$ is the saddle-point. The inequalities in \eqref{eq.M.saddle point} follow from \eqref{eq.M.T1.1}, cf. also \cite{Boyd04ConvexOptimization,Zeidler86NonlinearFunctionalAnalysisI}. It is remarkable that this result holds for any $\bW_1$ and hence does not require the channel to be degraded (unlike all known to date results). The saddle-point property in Proposition \ref{prop.M2} is instrumental in establishing the optimality of Gaussian signaling and hence the compound secrecy capacity for the non-degraded case (using this property avoids the need to prove the converse directly - the most difficult part of establishing the compound secrecy capacity for the non-degraded case).

The inequalities in \eqref{eq.M.saddle point} have the well-known game-theoretic interpretation: the transmitter sets $\bR=\bR^*$ and the adversary (nature or eavesdropper) sets $\bW_2=\epsilon \bI$; neither player can deviate from this strategy without incurring a penalty (provided that the other player follows it).

Note that the optimal signaling directions that achieve the compound capacity are the same as those for the regular MIMO channel (no eavesdropper) but the power allocation $\{\lambda_i^*\}$ is somewhat different from the regular water-filling (WF), even though it shares many of its properties, which is summarized below (see \cite{Loyka13FurtherResultsSecureMIMO} for further details).

\begin{proposition}
\label{prop.M.WF.prop}
Properties of the optimum power allocation:
\begin{enumerate}
	\item $\lambda_i^*$ is an increasing function of $g_i$ (strictly increasing unless $\lambda_i^*=0$ or $P_T$) , i.e. stronger eigenmodes get more power (as in the standard WF).

	\item $\lambda_i^*$ is an increasing function of $P_T$ (strictly increasing unless $\lambda_i^*=0$). $\lambda_i^* =0$ for $i > 1$ and $\lambda_1^* = P_T$ as $P_T \rightarrow 0$ if $g_1 > g_2$, i.e. only the strongest eigenmode is active at low SNR, and $\lambda_i^* >0$ if $g_i > \epsilon$ as $P_T \rightarrow \infty$, i.e. all sufficiently strong eigenmodes are active at high SNR.

	\item $\lambda_i^* > 0$ only if $g_i > \epsilon$, i.e. only the legitimate eigenmodes stronger than the eavesdropper ones can be active.

	\item $\lambda$ is a strictly decreasing function of $P_T$ and  $0<\lambda < g_1 - \epsilon$; $\lambda \rightarrow 0$ as $P_T \rightarrow \infty$ and $\lambda \rightarrow g_1 - \epsilon$ as $P_T \rightarrow 0$.

	\item There are $m_+$ active eigenmodes if the following inequalities hold:
\begin{eqnarray}
P_{m_+} < P_T \le P_{m_+ +1}
\end{eqnarray}
where $P_{m_+}$ is a threshold power (to have at least $m_+$ active eigenmodes):
\begin{align}
\label{eq.M.m+}
P_{m_+} \!\!=\!\! \sum_{i=1}^{m_+ -1} \!\frac{\epsilon + g_i}{2\epsilon g_i} \left(\!\sqrt{1+ \frac{4\epsilon g_i}{(\epsilon + g_i)^2} \frac{g_i - g_{m_+}}{(g_{m_+} - \epsilon)_+}} -1\! \right)\!,
\end{align}
for $m_+ = 2,...,N_T$ and $P_1 = 0$, so that $m_+$ is an increasing function of $P_T$. \end{enumerate}
\end{proposition}

The two terms in \eqref{eq.M.T1.4} represent the high-SNR asymptote and its (negative) correction term of the secrecy capacity respectively, so that
\bal
\label{eq.M.high SNR}
C^*(\epsilon) \rightarrow \sum_{i+} \ln \frac{g_i}{\epsilon}, \ i_+:g_{i+} > \epsilon,
\eal
as $SNR\rightarrow\infty$. In this regime, only those eigenmodes are active which are stronger than the eavesdropper ($g_{i+} > \epsilon$). Since the 2nd term is negative and increasing, it follows that
\bal
C^*(\epsilon) \leq \sum_{i+} \ln \frac{g_i}{\epsilon}, \ i_+:g_{i+} > \epsilon,
\eal
at any SNR. Fig. \ref{fig:Fig_1} illustrates this regime.

At low SNR, only the strongest mode is active and
\bal
C^*(\epsilon) = \ln \frac{1 + g_1 P_T}{1 + \epsilon P_T} \approx (g_1-\epsilon)P_T
\eal
where $g_i$ are in decreasing order, and 2nd equality holds when $(g_1-\epsilon)P_T \ll 1$. It follows from \eqref{eq.M.m+} that only one eigenmode is active, i.e. beamforming is optimal (which is appealing from practical perspective due to its low complexity), when
\begin{align}
\label{eq.M.beamforming}
P_T \leq \frac{\epsilon + g_1}{2\epsilon g_1} \left(\sqrt{1+ \frac{4\epsilon g_1}{(\epsilon + g_1)^2} \frac{g_1 - g_2}{(g_2 - \epsilon)_+}} -1 \right)
\end{align}
In particular, it is the case at any SNR if $g_2 \leq \epsilon$ (provided that $g_1 > \epsilon$), i.e. when the eavesdropper uncertainty is sufficiently large.

\subsection{Broader Class of Compound MIMO Channels}
\label{sec:uncert_broadclass}

\begin{figure}
  \centering
  	\scalebox{0.75}{\begin{picture}(0,0)%
  	\includegraphics{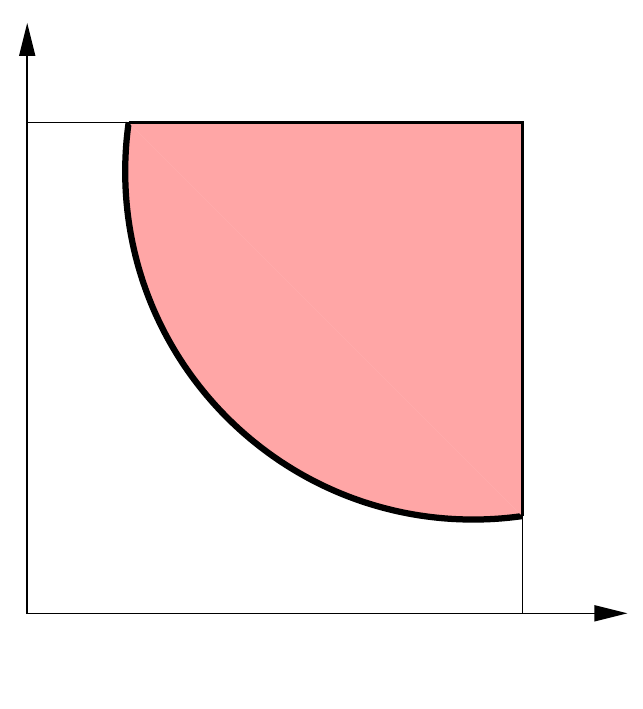}%
  	\end{picture}%
  	\setlength{\unitlength}{3947sp}%
  	\begingroup\makeatletter\ifx\SetFigFont\undefined%
  	\gdef\SetFigFont#1#2#3#4#5{%
  	  \fontfamily{#3}\fontseries{#4}\fontshape{#5}%
  	  \selectfont}%
  	\fi\endgroup%
  	\begin{picture}(3082,3393)(2375,-3491)
  	\put(2390,-729){\makebox(0,0)[rb]{\smash{{\SetFigFont{20}{24.0}{\rmdefault}{\mddefault}{\updefault}{\color[rgb]{0,0,0}$\epsilon$}%
  	}}}}
  	\put(5442,-3386){\makebox(0,0)[rb]{\smash{{\SetFigFont{20}{24.0}{\rmdefault}{\mddefault}{\updefault}{\color[rgb]{0,0,0}$d_1$}%
  	}}}}
  	\put(2395,-356){\makebox(0,0)[rb]{\smash{{\SetFigFont{20}{24.0}{\rmdefault}{\mddefault}{\updefault}{\color[rgb]{0,0,0}$d_2$}%
  	}}}}
  	\put(4889,-3347){\makebox(0,0)[b]{\smash{{\SetFigFont{20}{24.0}{\rmdefault}{\mddefault}{\updefault}{\color[rgb]{0,0,0}$\epsilon$}%
  	}}}}
  	\put(3901,-1486){\makebox(0,0)[lb]{\smash{{\SetFigFont{20}{24.0}{\rmdefault}{\mddefault}{\updefault}{\color[rgb]{0,0,0}$\boldsymbol{\mathcal{S}}_b$}%
  	}}}}
  	\put(2939,-2636){\makebox(0,0)[lb]{\smash{{\SetFigFont{20}{24.0}{\rmdefault}{\mddefault}{\updefault}{\color[rgb]{0,0,0}$\boldsymbol{\mathcal{S}}_a$}%
  	}}}}
  	\end{picture}%
  	}
  \caption{An example of two uncertainty sets when $\bW_2=\diag\{d_1,d_2\}\geq0$. The (whole) set $\vec{\sS}_a$ corresponds to the uncertainty set given in \eqref{eq.M3}, while the shaded set $\vec{\sS}_b$ corresponds to \eqref{eq.prop.M3}.}
  \label{fig1}
\end{figure}
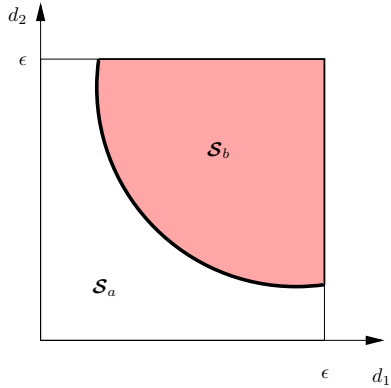

The result in Theorem \ref{thm.M1} can be further extended to a broader class of compound MIMO channels. To this end, let us generalize the uncertainty set $\sSt$ for the eavesdropper channel as follows
\begin{equation}
\label{eq.prop.M3}
	\bW_2 \in \sSt \rightarrow \bW_2 \le \epsilon \bI \in \sSt,
\end{equation}
i.e., all its members are less than or equal to $\epsilon \bI$. Unlike \eqref{eq.M3}, it may include \textit{not all} such $\bW_2$; it is not required to be convex, compact etc. Fig. \ref{fig1} illustrates the difference between the uncertainty sets defined in \eqref{eq.M3} and \eqref{eq.prop.M3} for diagonal $\bW_2$.

\begin{proposition}
\label{prop.M3}
Consider the compound Gaussian MIMO wiretap channel in \eqref{eq.M1} when $\bW_1$ is known and unknown $\bW_2$ belongs to the uncertainty set $\sSt$ in \eqref{eq.prop.M3}. Its compound secrecy capacity is $C_c=C^\ast(\epsilon)$, i.e., as in Theorem \ref{thm.M1}.
\end{proposition}
\begin{IEEEproof}
Observe that the compound secrecy capacity of this channel is not smaller than that in Theorem \ref{thm.M1}, since the uncertainty set here is included in the uncertainty set of Theorem \ref{thm.M1} (which includes \textit{all} $\bW_2 \le \epsilon \bI$, since it is equivalent to  $\lambda_1(\bW_2) \le \epsilon$). On the other hand, setting $\bW_2 = \epsilon \bI$ demonstrates that the lower bound is achieved by this worst-case channel. Since the compound capacity does not exceed the worst-case one, the desired result follows.
\end{IEEEproof}
\vspace*{0.5\baselineskip}

We remark that the set $\sSt$ is not necessarily convex or compact (as required by Theorem \ref{the:continuous}), nor it has some other ``nice'' properties, except that $\epsilon \bI$ is its dominant element, and that Theorem \ref{thm.M1} is a special case. This clearly demonstrates the importance of the isotropic eavesdropper for compound MIMO wiretap channels.

To generalize these results further, we will need the following definitions.

\begin{definition}
\label{def:uniquemaximum}
Let $\sSt$ be an uncertainty set of $\bW_2$. $\bW_2^*$ is its (unique) maximum element if $\bW_2^* \in \sSt$ and $\forall \bW_2 \in \sSt \rightarrow \bW_2 \le \bW_2^*$.
\end{definition}

\begin{definition}
\label{def:maximalelement}
$\bW_{2m}$ is a maximal element of $\sSt$ if $\bW_{2}, \bW_{2m} \in \sSt,   \bW_2 \ge \bW_{2m} \rightarrow \bW_2 = \bW_{2m}$ (i.e. the only element in $\sSt$ greater or equal to $\bW_{2m}$ is $\bW_{2m}$ itself).
\end{definition}

\begin{figure}
  \centering
  	\scalebox{0.75}{\begin{picture}(0,0)%
  	\includegraphics{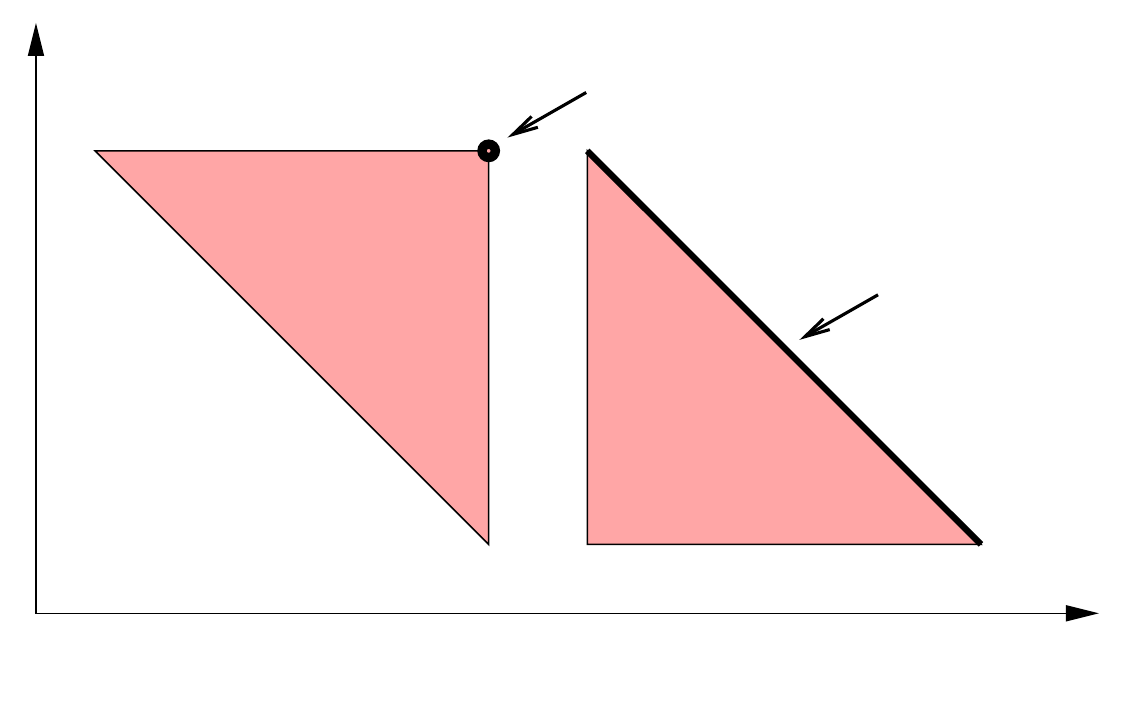}%
  	\end{picture}%
  	\setlength{\unitlength}{3947sp}%
  	\begingroup\makeatletter\ifx\SetFigFont\undefined%
  	\gdef\SetFigFont#1#2#3#4#5{%
  	  \fontfamily{#3}\fontseries{#4}\fontshape{#5}%
  	  \selectfont}%
  	\fi\endgroup%
  	\begin{picture}(5431,3393)(2380,-3491)
  	\put(3997,-1429){\makebox(0,0)[lb]{\smash{{\SetFigFont{20}{24.0}{\rmdefault}{\mddefault}{\updefault}{\color[rgb]{0,0,0}$\boldsymbol{\mathcal{S}}_a$}%
  	}}}}
  	\put(7796,-3386){\makebox(0,0)[rb]{\smash{{\SetFigFont{20}{24.0}{\rmdefault}{\mddefault}{\updefault}{\color[rgb]{0,0,0}$d_1$}%
  	}}}}
  	\put(2395,-356){\makebox(0,0)[rb]{\smash{{\SetFigFont{20}{24.0}{\rmdefault}{\mddefault}{\updefault}{\color[rgb]{0,0,0}$d_2$}%
  	}}}}
  	\put(5023,-486){\makebox(0,0)[lb]{\smash{{\SetFigFont{20}{24.0}{\rmdefault}{\mddefault}{\updefault}{\color[rgb]{0,0,0}$\boldsymbol{W}_2^*$}%
  	}}}}
  	\put(6459,-1449){\makebox(0,0)[lb]{\smash{{\SetFigFont{20}{24.0}{\rmdefault}{\mddefault}{\updefault}{\color[rgb]{0,0,0}$\boldsymbol{\mathcal{S}}_{2m}$}%
  	}}}}
  	\put(5670,-2279){\makebox(0,0)[lb]{\smash{{\SetFigFont{20}{24.0}{\rmdefault}{\mddefault}{\updefault}{\color[rgb]{0,0,0}$\boldsymbol{\mathcal{S}}_b$}%
  	}}}}
  	\end{picture}%
  	}
  \caption{An example of two uncertainty sets $\vec{\sS}_a$ and $\vec{\sS}_b$ when $m=2$ and $\bW_2 = \diag\{d_1,d_2\} \ge 0$. $\vec{\sS}_a$ has a (unique) maximum element $\bW_2^*$ (dark dot) while $\vec{\sS}_b$ does not, but only a set of maximal elements (dark line) $\vec{\sS}_{2m}$.}
  \label{fig2}
\end{figure}
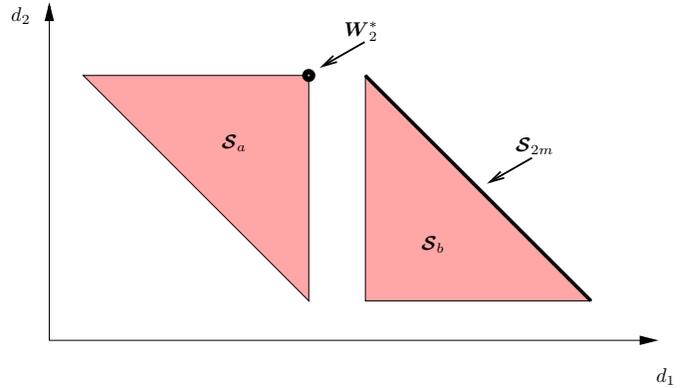

Note that Definition \ref{def:maximalelement} is due to the fact that not any two positive semi-definite matrices can be compared (i.e. it can be that neither $\bW_1 \ge \bW_2$ nor $\bW_1 < \bW_2$ is true, unlike the scalar case), so that a maximum element may not exist. While maximum element, if it exists, is unique, there may be many maximal elements in a set (see e.g. \cite{Boyd04ConvexOptimization} for more details). Fig.~\ref{fig2} illustrates these definitions for the case of diagonal $\bW_2$ and $m=2$.

We are now in a position to generalize Proposition \ref{prop.M3}.

\begin{proposition}
\label{prop.M4}
Consider the compound Gaussian MIMO wiretap channel in \eqref{eq.M1} when $\bW_1$ is known and unknown $\bW_2$ belongs to an uncertainty set $\sSt$, whose maximum element is $\bW_2^*$. The saddle-point property holds, so that the compound secrecy capacity $C_c$ equals to the worst-case secrecy capacity $C_w$:
\begin{align} \notag
\label{eq.M.P4.1}
C_c &= \mathop {\max }\limits_{\bR} \min_{\bW_2 \in \sSt} C(\bR,\bW_2) \\
\notag
&= \min_{\bW_2 \in \sSt} \mathop {\max }\limits_{\bR} C(\bR,\bW_2) = C_w\\
&= \mathop{\max }\limits_{\bR} C(\bR,\bW_2^*)
\end{align}
where the worst-case channel is $\bW_2^*$, and the transmission on this channel is optimal for the whole class of channels in $\sSt$.
\end{proposition}
\begin{IEEEproof}
Observe that
\begin{align}
\label{eq.M.P4.2}
C(\bR,\bW_2) \ge C(\bR,\bW_2^*) \ \forall \bR, \bW_2 \in \sSt
\end{align}
which is due to the fact that $|\bI+\bW \bR|$ is monotonically increasing in $\bW$ \cite{Zhang11MatrixTheory} for any (positive semi-definite) $\bR$, so that, by using $\max\min$ and $\min\max$ on both sides, one obtains
\begin{align} \notag
\mathop {\max }\limits_{\bR} \min_{\bW_2 \in \sSt} C(\bR,\bW_2) &= \mathop {\max }\limits_{\bR} C(\bR,\bW_2^*) \\
\label{eq.M.P4.3}
&= \min_{\bW_2 \in \sSt} \mathop {\max }\limits_{\bR} C(\bR,\bW_2) = C_w.
\end{align}
To prove the operational meaning of the $\max \min$ part, observe that Corollary 2 does not apply directly as $\sSt$ is not necessarily compact. Instead, consider another compact set $\sSt'$ that includes \textit{all} positive semi-definite $\bW_2$ such that $\bW_2 \le \bW_2^*$. Clearly, this set is closed and bounded and hence compact, and $\sSt \subseteq \sSt'$, so that its compound capacity $C_c'$ satisfies $C_c' \le C_c$. Applying Corollary 2 to $\sSt'$, one obtains
\begin{align} \notag
C_w &= \min_{\bW_2 \in \sSt} \mathop {\max }\limits_{\bR} C(\bR,\bW_2)\\ \notag
&= \min_{\bW_2 \in \sSt'} \mathop {\max }\limits_{\bR} C(\bR,\bW_2)\\ \notag
&= \mathop {\max }\limits_{\bR} \min_{\bW_2 \in \sSt'} C(\bR,\bW_2) \\ \notag
&\le C_c' \le C_c \le C_w
\end{align}
where the 2nd equality is due to the fact that \eqref{eq.M.P4.2} holds for $\sSt'$ as well so that $C_w$ is the same for $\sSt$ and $\sSt'$ (since both sets have the same maximum element $\bW_2^*$); the 3rd equality is due to the fact that \eqref{eq.M.P4.3} holds for $\sSt'$ as well. This proves $C_c'=C_c=C_w$ and hence the desired result.
\end{IEEEproof}
\vspace*{0.5\baselineskip}

This proposition says, in effect, that the saddle-point property holds and, thus, the compound secrecy capacity equals the worst-case one, if a maximum element of the uncertainty set $\sSt$ exists\footnote{Recall that it is not the case in general and many sets of positive semi-definite matrices do not have a maximum element, as Fig. \ref{fig2} shows.} and the rest of its structure is irrelevant.

When the uncertainty set does not have a maximum element, its compound and worst-case secrecy capacities can be characterized using minimal elements as follows.

\begin{proposition}
\label{prop.M5}
Consider the compound Gaussian MIMO channel in \eqref{eq.M1} when $\bW_1$ is known and unknown $\bW_2$ belongs to a bounded and closed uncertainty set $\sSt$, which does not have a maximum element. Then,
\begin{align}
\label{eq.M.P5.1}
\min_{\bW_2 \in \sSt} C(\bR,\bW_2) = \min_{\bW_2 \in \vec{\sS}_{2m}} C(\bR,\bW_2) \ \forall \bR
\end{align}
where $\vec{\sS}_{2m}$ is the set of all maximal elements $\bW_{2m}$ of $\sSt$, and hence
\begin{align}
\label{eq.M.P5.2}
C_w &= \min_{\bW_2 \in \sSt} \mathop {\max }\limits_{\bR} C(\bR,\bW_2) = \min_{\bW_2 \in \vec{\sS}_{2m}} \mathop {\max }\limits_{\bR} C(\bR,\bW_2)\\
C_c &\ge \max_{\bR} \min_{\bW_2 \in \sSt} C(\bR,\bW_2) = \max_{\bR} \min_{\bW_2 \in \vec{\sS}_{2m}} C(\bR,\bW_2)
\end{align}
i.e. minimizing over the whole uncertainty set $\sSt$ is equivalent to minimizing over (normally much smaller) set of its maximal elements.
\end{proposition}
\begin{IEEEproof}
Since the proof is highly technical, it is relegated to Appendix \ref{app:prop.M5}.
\end{IEEEproof}
\vspace*{0.5\baselineskip}

We remark that Proposition \ref{prop.M5} effectively reduces the dimensionality of the related optimization problem: if the original problem in \eqref{eq.M.P5.2} is $D$-dimensional, the reduced one (on the right hand side) is at most $(D-1)$-dimensional, since $\vec{\sS}_{2m}$ is on the boundary of $\sSt$ (this can be proved by contradiction). In some cases, this proposition can be applied even if $\sSt$ is not compact by enclosing it in a bigger compact set $\sSt'$ provided that the minimum in \eqref{eq.M.P5.1} is the same for both sets.

The last two propositions demonstrate the key role of the maximum element in the uncertainty set: if it exists, a saddle-point exists, so it is a sufficient condition. It can be shown, via examples, that the absence of a maximum element may or may not result in the absence of a saddle-point, so there is no necessary condition here.

\subsection{Rank-Constrained Eavesdropper}

In this section, we consider the case where there is an extra constraint on the rank $r(\bW_2)$ of the eavesdropper channel $\bW_2$, $r(\bW_2) \le r_2$ for given $r_2 \le N_T$. This constraint is motivated by the fact that $r(\bW_2) \le N_2$ so that when the number $N_2$ of eavesdropper antennas is small, $N_2 \le N_T$, full-rank $\bW_2$ is not possible so that the results in Theorem 4 may be too conservative\footnote{This problem formulation was suggested by A. Khisti.}. This applies in particular to a massive MIMO case, where the transmitter is a base station with a large number of antennas and the receiver/eavesdropper are handsets with a small number of antennas (due to the size/complexity constraints), so that $N_T \gg N_1, N_2$.

The eavesdropper uncertainty set is of the form
\begin{align}
\label{eq.M.r1}
	\vec{\sS}_{2a} =\big\{\bW_2:|\bW_2|_2 \le \epsilon,\ r(\bW_2) \le r_2 \big\}
\end{align}
where the 1st inequality reflects the fact that the eavesdropper channel gain is bounded (due to e.g. minimum propagation path loss) and the 2nd one reflects the fact that the rank is bounded due to e.g. small number of eavesdropper antennas. The compound secrecy capacity can now be characterized as follows.

\begin{theorem}
\label{thm.M.r1}
Consider the compound Gaussian MIMO wiretap channel in \eqref{eq.M1} with known $\bW_1$ and unknown $\bW_2$ belonging to the uncertainty set $\vec{\sS}_{2a}$ in \eqref{eq.M.r1}; assume that the rank of the legitimate channel satisfies $r(\bW_1)=r_1 \le r_2$. The compound secrecy capacity $C_c$ of this channel is as follows:
\begin{align} \notag
\label{eq.M.r.T1.1}
C_c &= \mathop {\max }\limits_{\bR} \min_{\bW_2} C(\bR,\bW_2) \\
\notag
&= \min_{\bW_2} \mathop {\max }\limits_{\bR} C(\bR,\bW_2) \\
&= \sum_{i=1}^{r_1} \ln \frac{1 + g_i \lambda_i^*}{1 + \epsilon \lambda_i^*} = C^*(\epsilon)
\end{align}
where $\max$ and $\min$ are over all admissible $\bR, \bW_2$:
$ \bR, \bW_2 \ge 0$, $\tr \bR \le P_T$, $\bW_2 \in \vec{\sS}_{2a}$. The optimal signaling is Gaussian and on the eigenmodes of the legitimate channel as in \eqref{eq.M.T1.2}, and $\lambda_i^*$ is as in \eqref{eq.M.T1.3}. The worst-case eavesdropper is $\bW_2^*=\epsilon \bU_{1a}\bU_{1a}^+$, where the columns of semi-unitary matrix $\bU_{1a}$ are the eigenvectors of $\bW_1$ corresponding to strictly positive eigenvalues.
\end{theorem}

\begin{IEEEproof}
First, observe that $\{\sigma_i(\bH\bR^{\frac{1}{2}})\}$ is weakly majorized by $\{\sigma_i(\bH)\sigma_i(\bR^{\frac{1}{2}})\}$ (see e.g. \cite{HornJohnson91TopicsMatrixAnalysis}), i.e.
\bal
\sum_{i=1}^k \sigma_i(\bH\bR^{\frac{1}{2}}) \le \sum_{i=1}^k \sigma_i(\bH)\sigma_i(\bR^{\frac{1}{2}}),\ 1 \le k \le N_T
\eal
where all singular values $\sigma_i$ are in decreasing order. Therefore,
\begin{align} \notag
\label{eq.M.r.T1.2}
\ln |\bI+\bW_2\bR| &= \sum_{i=1}^{r_2} \ln(1+\sigma_i^2(\bH_2\bR^{1/2})) \\ \notag
&\le \sum_{i=1}^{r_2} \ln(1+\sigma_i^2(\bH_2)\sigma_i^2(\bR^{1/2})) \\
&\le \sum_{i=1}^{r_2} \ln(1+ \epsilon \lambda_i(\bR))
\end{align}
where we have used the fact that $\sigma_i^2(\bR^{\frac{1}{2}})=\lambda_i(\bR)$, $\sigma_i^2(\bH)=\lambda_i(\bW)$. The 1st inequality is due to \cite[Theorem 3.3.14]{HornJohnson91TopicsMatrixAnalysis} and the fact that $\ln(1+e^x)$ is convex in $x$ and $\ln(1+x^2)$ is continuous, and the 2nd inequality is due to $\lambda_i(\bW_2) \le |\bW_2|_2 \le \epsilon$. Similarly, we have
\begin{align}
\label{eq.M.r.T1.3}
\ln |\bI+\bW_1\bR| \le \sum_{i=1}^{r_1} \ln(1+\lambda_i(\bW_1)\lambda_i(\bR)).
\end{align}
Using these two upper bounds and observing that the 2nd one is achieved by using $\bR$ with the same eigenvectors as those of $\bW_1$ and such choice of eigenvectors does not affect the bound in \eqref{eq.M.r.T1.2}, one obtains, using Theorem 3:
\begin{align} \notag
\label{eq.M.r.T1.4}
C_c &\ge \max_{\bR} \min_{\bW_2} C(\bR,\bW_2)\\ \notag
 &\ge \max_{\bR}\{ \ln |\bI+\bW_1\bR| - \sum_{i=1}^{r_2} \ln(1+ \epsilon\lambda_i(\bR))\}\\
 &= \max_{\lambda_i:\ \lambda_i \ge 0, \sum_i \lambda_i \le P_T} \sum_{i=1}^{r_1} \ln\frac{1+\lambda_i(\bW_1)\lambda_i}{1+ \epsilon \lambda_i} = C^*(\epsilon)
\end{align}
where the sum is limited to $r_1$ due to $r_1 \le r_2$ so that, from Corollary 1 in \cite{Loyka12OptimalSignalingMIMOWiretap}, $r(\bR^*)\le r_1$. On the other hand, since the worst-case capacity is not less than the compound one,
\begin{align} \notag
\label{eq.M.r.T1.5}
C_c \le C_w &= \min_{\bW_2} \max_{\bR} C(\bR,\bW_2)\\ \notag
 &\le \max_{\bR} C(\bR,\bW_2^*)\\
 &= \max_{\lambda_i:\ \lambda_i \ge 0, \sum_i \lambda_i \le P_T} \sum_{i=1}^{r_1} \ln\frac{1+\lambda_i(\bW_1)\lambda_i}{1+ \epsilon \lambda_i} = C^*(\epsilon)
\end{align}
where the 2nd equality is due to the fact that when $\bW_1$ and $\bW_2$ have the same eigenvectors, signaling on those eigenvectors is optimal (see \cite[Proposition 1]{Loyka14RankDeficientMIMOWiretap}). This proves the saddle-point and thus establishes the capacity $C_c=C_w=C^*(\epsilon)$. The optimal signaling follows from \eqref{eq.M.r.T1.2} and \eqref{eq.M.r.T1.3} where the equalities are attained by $\bW_2=\epsilon \bU_{1a}\bU_{1a}^+$ and $\bR=\bU_{1a}\bL\bU_{1a}^+$, which also attains the equalities in \eqref{eq.M.r.T1.4} and \eqref{eq.M.r.T1.5}.
\end{IEEEproof}

\begin{remark}
Note that the worst-case eavesdropper $\bW_2^*=\epsilon \bU_{1a}\bU_{1a}^+$ is "isotropic" on the sub-space spanned by the columns of $\bU_{1a}$ (but not on the whole space), which is known as ``omni-directional'' in the antenna literature \cite{Balanis05AntennaTheory} (i.e. having the same gain in all directions of that sub-space). Comparing Theorems 4 and 5, one concludes that the eavesdropper rank constraint has no effect on the capacity and optimal signaling provided that $r_1 \le r_2$ holds.
\end{remark}

\begin{remark}
Unlike the rank-unconstrained case, there is no dominant channel in the rank-constrained uncertainty set, i.e. $\bW_2 \le \bW_2^*$ does not hold for all $\bW_2 \in \vec{\sS}_{2a}$, so that the uncertainty set is not "degraded" (with respect to $\bW_2^*$ or any other $\bW_2$).
Since the set $\vec{\sS}_{2a}$ is not convex either, one cannot use Von Neumann mini-max Theorem (or its extensions) to infer an existence of saddle-point, which is established in \eqref{eq.M.r.T1.1} via the singular value inequalities, so that the following inequalities hold for any feasible $\bR$ and $\bW_2$,
\begin{align}
\label{eq.M.saddle point.r}
C(\bR, \bW_2^*) \le C_c = C_w= C(\bR^*,\bW_2^*) \le C(\bR^*,\bW_2)
\end{align}
where $(\bR^*, \bW_2^*)$ is the saddle-point. It can be demonstrated (via examples) that the saddle-point property does not hold if $r_1 > r_2$.
\end{remark}

\begin{remark}
The condition on the ranks $r_1 \le r_2$ is insured if $N_1 \le N_2$ and both channels are of full raw ranks. In particular, this holds if $N_1=N_2=1$.
\end{remark}

\section{Double-Sided Channel Uncertainty}
\label{sec:double-sided}

Here we consider the case where both the legitimate and eavesdropper channels are uncertain. The compound channel model follows the model in \eqref{eq.M1} where:
\begin{subequations}
\label{eq.M4}
\begin{align}
	\sSo&= \big\{\bH_1: \bH_1 = \bH_{0} + \Delta\bH,\ |\Delta\bH|_2 \le \epsilon_1\big\} \\
	\sSt&= \big\{\bW_2: |\bW_2|_2 \le \epsilon\big\}
\end{align}
\end{subequations}
where $\bH_{0}$ is the nominal part of $\bH_1$ known to the transmitter, and $ \Delta\bH$ is the uncertain, unknown part; $|\Delta\bH|_2=\sigma_1(\Delta\bH)$ is the spectral norm of $\Delta\bH$, i.e. the largest singular value $\sigma_1(\Delta\bH)$. The uncertainty of $\bW_2$ follows the same model as in \eqref{eq.M3}. This compound model reflects two important points:

\begin{enumerate}
\item The desire of the eavesdropper to be confidential to keep its spying abilities uncompromised, so it does not share its channel with the transmitter and therefore only minimal information about $\bH_2$ is available to the latter.
\item The legitimate receiver, on the other hand, wishes to maximize the rate so it shares its channel with the transmitter. Its channel uncertainty is due to the limitations of the feedback and estimation procedure, which is normally much smaller than that of the eavesdropper (and hence the known nominal part).
\end{enumerate}

The secrecy capacity of this compound channel can be characterized as follows. For this purpose we define
\begin{equation*}
	C(\bR,\bW_1,\bW_2) = \ln\frac{|\bI+\bW_1\bR|}{|\bI+\bW_2\bR|}
\end{equation*}
which depends on the transmit covariance matrix $\bR$ and the unknown channels $\bW_1=\bH_1^+\bH_1$ and $\bW_2=\bH_2^+\bH_2$ to the legitimate receiver and the eavesdropper respectively.

\begin{theorem}
\label{thm.M2}
Consider the compound Gaussian MIMO wiretap channel in \eqref{eq.M1} when $\bW_1$ and $\bW_2$ are unknown and belong to the uncertainty sets $\sSo$ and $\sSt$ in \eqref{eq.M4}. Then, the compound secrecy capacity $C_c$ is
\begin{align}\notag
\label{eq.M.T2.1}
C_c &= \mathop {\max }\limits_{\bR} \min_{\bW_1, \bW_2} C(\bW_1, \bW_2, \bR) \\
\notag
&= \min_{\bW_1, \bW_2} \mathop {\max }\limits_{\bR} C(\bW_1, \bW_2, \bR) =C_w\\
&= C(\bW_{1w}, \bW_{2w}, \bR^*),
\end{align}
i.e., the worst-case secrecy capacity $C_w$ is also the (compound) secrecy capacity $C_c$ of the class of channels and Gaussian signaling is optimal. The saddle-point property holds,
\begin{align}\notag
\label{eq.M.T2.1a}
C(\bW_{1w}, \bW_{2w}, \bR) &\le C_c = C(\bW_{1w}, \bW_{2w}, \bR^*) \\
&\le C(\bW_{1}, \bW_2, \bR^*),
\end{align}
where $(\bW_{1w}, \bW_{2w}, \bR^*)$ is the saddle-point. The worst-case channel is
\begin{align}
\label{eq.M.T2.2} \notag
\bW_{1w} &= \bH_{1w}^+\bH_{1w}, \ \bH_{1w} = \bV_0(\bSg_0 - \epsilon_1\bI)_+\bU_0^+, \\
\bW_{2w} &= \epsilon\bI,
\end{align}
where $\bU_0, \bV_0$ are unitary matrices of right and left singular vectors of the nominal channel $\bH_{0}$ and $\bSg_0$ is the diagonal matrix of its singular values. The optimal covariance $\bR^*$ is as in Theorem \ref{thm.M1} with the substitution
\begin{align}
\label{eq.M.T2.3}
g_i \rightarrow (\sigma_i(\bH_0)-\epsilon_1)^2_+, \ \bU_1 \rightarrow \bU_0,
\end{align}
i.e., the optimal signaling is on the eigenmodes of the degraded nominal channel $\bH_{1w}$ and isotropic eavesdropper.
\end{theorem}
\begin{IEEEproof}
The proof can be found in Appendix \ref{app:thm.M2}.
\end{IEEEproof}
\vspace*{0.5\baselineskip}

Note that this theorem does not require the compound channel to be degraded. Remarkably, the saddle-point property still holds and the isotropic eavesdropper (of the maximum gain) is still the worst-case one, even under the legitimate channel uncertainty, and the optimal signaling is almost the same as in Theorem \ref{thm.M1} (Gaussian signaling is still optimal), with the legitimate channel substituted by its degraded (due to uncertainty) version. We observe that, as the uncertainty (i.e. $\epsilon_1$ and/or $\epsilon$) increases, fewer and fewer eigenmodes are used until only the strongest one remains active, in which case the beamforming is optimal (see \eqref{eq.M.beamforming}). From this perspective, beamforming is the most robust strategy.

The game-theoretic interpretation of the inequalities in \eqref{eq.M.T2.1a} is the same as for the single-sided uncertainty: $\{\bW_{1w}, \epsilon\bI, \bR^*\}$ is a saddle-point in the matrix game between the transmitter on one side and the eavesdropper and nature on the other; neither can deviate from the optimal strategy without incurring penalty provided that the other player follows the strategy.

\subsection{Rank-Constrained Eavesdropper}
Using similar arguments, Theorem \ref{thm.M2} can be extended to the rank-constrained eavesdropper channel,
\begin{subequations}
\label{eq.M6.2.1}
\begin{align}
	&\sSo = \big\{\bH_1: \bH_1 = \bH_{0} + \Delta\bH,\ |\Delta\bH|_2 \le \epsilon_1\big\} \\
	&\bsS_{2} = \big\{\bH_2: |\bH_2|_2 \le \epsilon,\ r(\bH_2) \le r_2  \big\}
\end{align}
\end{subequations}
where the eavesdropper rank is constrained by $r_2$ (due to e.g. limited number of antennas).

\begin{theorem}
\label{thm.M6.2}
Consider the compound Gaussian MIMO wiretap channel in \eqref{eq.M1} when $\bH_1$ and $\bH_2$ are unknown and belong to the uncertainty sets $\sSo$ and $\bsS_{2}$ in \eqref{eq.M6.2.1}. Assume that $r(\bH_0)=r_1 \le r_2$. Then, the compound secrecy capacity $C_c$ is as in \eqref{eq.M.T2.1}; the saddle-point property in \eqref{eq.M.T2.1a} holds and the worst-case channel $\bW_{1w}$ is as in \eqref{eq.M.T2.2} while $\bW_{2w}$ is
\begin{align}
\label{eq.M6.T2.1}
\bW_{2w} &= \epsilon^2 \bU_{0a}\bU_{0a}^+,\ \bH_{2w} = \bV \bSg_{2w} \bU_{0}^+,
\end{align}
where $\bV$ is an arbitrary unitary matrix, semi-unitary matrix $\bU_{0a}$ collects the columns of $\bU_0$ corresponding to strictly positive singular values, and
\bal
 \bSg_{2w} = diag\{\epsilon,..,\epsilon,0,..,0\}
\eal
is a diagonal matrix with the 1st $r_1$ diagonal entries being $\epsilon$ and 0 otherwise. The optimal covariance $\bR^*$ is as in Theorem \ref{thm.M2}, i.e. the optimal signalling is Gaussian and on the eigenmodes of the worst-case legitimate channel $\bH_{1w}$.
\end{theorem}
\begin{IEEEproof}
The proof can be found in Appendix \ref{app:thm.M6.2}.
\end{IEEEproof}
\vspace*{0.5\baselineskip}

\section{Weak vs. Strong Secrecy}
\label{sec:weak vs. strong}

The results above have been established under the strong secrecy condition. It was demonstrated in \cite{Csiszar96SecrecyCapacity,Maurer00WeakToStrongSecrecy} that, for regular (single-state or known) channels, strong and weak secrecy capacities are the same. That result, however, does not immediately apply to the compound setting here. Nevertheless, it can be shown that the weak $C_c^{weak}$ and strong $C_c^{strong}$ compound secrecy capacities are the same,
\bal
\label{eq.Cweak=Cstrong}
C_c^{weak}= C_c^{strong}
\eal
if the saddle-point property holds under strong secrecy, i.e. $C_w = C_c^{strong}$. Indeed, under the saddle point property,
\bal
\label{eq.Cweak=Cstrong proof}
C_w = C_c^{strong} \leq C_c^{weak} \leq C_w
\eal
from which \eqref{eq.Cweak=Cstrong} follows, where we have used the fact that the worst-case capacity is the same under the strong and weak secrecies, and that the strong compound secrecy capacity is not larger than the weak one. In particular, the results in Theorems 4, 5 and Proposition 5 also hold under weak secrecy, so that one can go from weak to strong secrecy for free in the compound settings as well under the saddle-point property.

In fact, the chain argument in \eqref{eq.Cweak=Cstrong proof} has the following implications:
\begin{itemize}
\item the saddle point under strong secrecy ($C_w = C_c^{strong}$) implies a saddle point under weak secrecy ($C_w = C_c^{weak}$),
\item no saddle point under weak secrecy ($C_w > C_c^{weak}$) implies no saddle point under strong secrecy ($C_w > C_c^{strong}$).
\end{itemize}

\section{Conclusion}
\label{sec:conclusion}

The secrecy capacity of compound wiretap channels has been studied. First, the achievable strong secrecy rate of finite-state compound channels under finite alphabets in \cite{Bjelakovic13CompoundWiretap} was extended to arbitrary uncertainty sets (not necessarily countable or finite-state) and then to continuous input/output alphabets and arbitrary compact uncertainty sets. Based on this, the (strong) secrecy capacity of the compound Gaussian MIMO wiretap channel has been established under the spectral norm constraint on the eavesdropper channel. The channel is not required to be degraded. The optimal signaling as well as the secrecy capacity are given in a closed form. The saddle-point property has been shown to hold, so that the compound capacity equals to the worst-case one and signaling on the worst-case channel achieves the compound capacity. Isotropic eavesdropper is the worst-case one and signaling on the eigenmodes of the legitimate channel is optimal. The results are extended to non-isotropic uncertainty sets. It is shown that the existence of a maximum element in the uncertainty set is sufficient for a saddle-point to exist, so that compound capacity equals to the worst-case one and signaling on the worst-case channel achieves the capacity of the whole class of channels. Finally, these results are extended to include the legitimate channel uncertainty.

While the results above have been established under the total power constraint $tr\bR \le P_T$, using similar reasoning it can be shown that the same result holds under a general power constraint of the form $\bR \in \vec{\mathcal{S}_\bR}$, where $\vec{\mathcal{S}_\bR}$ is a unitary invariant set of positive semi-definite matrices, i.e. $\bR \in \vec{\mathcal{S}_\bR}$ implies $\bU\bR\bU^+ \in \vec{\mathcal{S}_\bR}$ for any unitary $\bU$. This constraint limits possible eigenvalues of $\bR$ but does not constrain in any way its eigenvectors. Special cases include the total and maximum per-eigenmode power constraints (either alone or in combination with each other).

\appendix
\label{sec:Appendix}

\subsection{Proof of Lemma \ref{lem:comp1}}
\label{app:lem.1}

It is known that the secrecy capacity of a wiretap channel depends only on its marginal channels and not on its joint probability distribution\footnote{In particular, two wiretap channels with different joint probability distributions will have the same secrecy capacity if they share the same marginal channel probabilities.}, cf. for instance from \cite[Lemma 2.1]{Liang09InformationTheoreticSecurity}. Therefore, it suffices to find good approximations $(\overline{W}_s,\overline{V}_s)$ for the marginals $(W_s,V_s)$ only, which simplifies the task significantly. To this end, using \cite[Lemma 4]{Blackwell59Compound} for both marginal channels, one obtains approximations that satisfy
\begin{gather*}
|W_s(y|x)-\overline{W}_s(y|x)|\leq |\sY|/L \leq |\sY||\sZ|/L, \\
|V_s(z|x)-\overline{V}_s(z|x)|\leq |\sZ|/L \leq |\sY||\sZ|/L, \\
W_s(y|x)\leq2^{\frac{2|\sY|^2}{L}}\overline{W}_s(y|x)\leq2^{\frac{2|\sY|^2|\sZ|^2}{L}}\overline{W}_s(y|x), \\
V_s(z|x)\leq2^{\frac{2|\sZ|^2}{L}}\overline{V}_s(z|x)\leq2^{\frac{2|\sY|^2|\sZ|^2}{L}}\overline{V}_s(z|x)
\end{gather*}
for all $x\in\sX$, $y\in\sY$, and $z\in\sZ$, and further for any input distribution $P_X\in\sP(\sX)$
\begin{align*}
|I(X;Y_s)-I(X;\overline{Y}_s)| &\leq 2|\sY|^{3/2}/L^{1/2}\leq 2(|\sY||\sZ|)^{3/2}/L^{1/2}, \\
|I(X;Z_s)-I(X;\overline{Z}_s)| &\leq 2|\sZ|^{3/2}/L^{1/2}\leq 2(|\sY||\sZ|)^{3/2}/L^{1/2}.
\end{align*}
Note that in the first step, the application of \cite[Lemma 4]{Blackwell59Compound} yields bounds, where the constants are different and depend on their own alphabet size, i.e., either on $|\sY|$ or on $|\sZ|$, which is difficult to use in the following analysis. The 2nd step  results in the bounds with the same constant, which facilitates the further analysis. \hfill\IEEEQED

\subsection{Proof of Lemma \ref{lem:comp2}}
\label{app:lem.2}

The first property \eqref{eq:comp2_error} follows by observing that for all $x^n\in\sX^n$ and $y^n\in\sY^n$ we have
\begin{align*}
	W_s^n(y^n|x^n)&=\prod_{i=1}^nW_s(y_i|x_i)\\
		&\leq 2^{n\frac{|\sY|^2|\sZ|^2}{L}}\prod_{i=1}^n\overline{W}_s(y_i|x_i)\\
		&=2^{n\frac{|\sY|^2|\sZ|^2}{L}}\overline{W}_s^n(y^n|x^n)
\end{align*}
which naturally extends to decoding sets as $W_s^n(\sD_m^c|x^n)\leq 2^{n\frac{|\sY|^2|\sZ|^2}{L}}\overline{W}_s^n(\sD_m^c|x^n)$ and likewise for the error probability.

The more interesting part is the robustness of the secrecy constraint. Following the classical approach in \cite[Lemma 4]{Blackwell59Compound} would lead to a bound which is too loose to prove what we aim for, cf. also Remark \ref{rem:continuity}. Therefore, we make use of a recent result in \cite[Lemma 2]{BocheSchaeferPoorXXContinuity}.

\begin{lemma}
\label{lem:continuity}
Let $\sX$ and $\sY$ be finite alphabets and $W,\Wtil:\sX\rightarrow\sP(\sY)$ be arbitrary channels with
\begin{equation}
	\max_{x\in\sX}\sum_{y\in\sY}|W(y|x)-\Wtil(y|x)| \leq \epsilon
	\label{eq:cc_lem2_d}
\end{equation}
for some $\epsilon>0$. For arbitrary $n\in\N$, let $\sU$ be an arbitrary finite set, $P_U\in\sP(\sU)$ the uniform distribution on $\sU$, and $E(x^n|u)$, $x^n\in\sX^n$ an arbitrary stochastic encoder, cf. \eqref{eq:dmc_encoder}. We consider the probability distributions
\begin{align*}
	P_{UY^n}(u,y^n) &= \sum_{x^n\in\sX^n}W^n(y^n|x^n)E(x^n|u)P_U(u) \\
	\Ptil_{UY^n}(u,y^n) &= \sum_{x^n\in\sX^n}\Wtil^n(y^n|x^n)E(x^n|u)P_U(u).
\end{align*}
Then it holds that
\begin{equation}
	\big|I(U;Y^n\|P) - I(U;Y^n\|\Ptil)\big| \leq 4 n \big(\epsilon\log|\sY| +H_2(\epsilon)\big)
	\label{eq:cc_lem2}
\end{equation}
where $I(U;Y^n\|P)$ means that the mutual information is evaluated under the joint probability distribution $P$.
\end{lemma}
\begin{IEEEproof}
The proof is based on the technique developed for quantum channels in \cite{Leung09ContinuityQuantumCapacities} and can be found in Appendix B of \cite{BocheSchaeferPoorXXContinuity}.
\end{IEEEproof}

Note that this lemma must be applied carefully: In the problem at hand, the channels $V_s$ and $\overline{V}_s$ satisfy $|V_s(z|x)-\overline{V}_s(z|x)|\leq|\sY||\sZ|/L$ for all $x\in\sX$ and $z\in\sZ$, cf. \eqref{eq:comp1_prob1}-\eqref{eq:comp1_prob2}. Thus, \eqref{eq:cc_lem2_d} is satisfied with $\epsilon=|\sY||\sZ|^2/L$ which then yields the desired result, i.e.,
\begin{equation}
	|I(M;Z_s^n)- I(M;\overline{Z}_s^n)|\leq 4n\big(|\sY||\sZ|^2\log|\sZ|/L+ H_2(|\sY||\sZ|^2/L)\big).
	\label{eq:app_comp2}
\end{equation}
This completes the proof. \hfill\IEEEQED

\subsection{Proof of Proposition \ref{prop.M5}}
\label{app:prop.M5}

The following lemma is instrumental.

\begin{lemma}
\label{lemma.M.convergence}
Let $\bW_1, \bW_2, ...$ be a bounded and increasing sequence of positive semi-definite matrices, i.e.
\begin{align}
\vec{0} \le \bW_1 \le \bW_2 \le .. \le \bW_i \le ... \le a\bI
\end{align}
where $0<a<\infty$ is a positive constant. This sequence converges.
\end{lemma}
\begin{IEEEproof}
Consider the following sequence of (non-negative) scalars $\alpha_i = \bx^+ \bW_i \bx$, where $\bx$ is a vector of appropriate size; for convenience, we take $|\bx|=1$. Since $\{\bW_i\}$ is an increasing and bounded sequence, so is $\{\alpha_i\}$,
\begin{align}
0 \le \alpha_1 \le \alpha_2 \le .. \le \alpha_i \le ... \le a
\end{align}
and therefore it converges to some non-negative number $b(\bx) = \lim_{i\rightarrow\infty} \alpha_i \le a$. Hence, for any $\epsilon >0$, there is such $n(\epsilon, \bx)$ that $b(\bx) - \alpha_i < \epsilon \ \forall i>n(\epsilon, \bx), \bx$. Since this is true for any $\bx$, take $n(\epsilon) = \max_{\bx} n(\epsilon, \bx)$ and observe that $|b(\bx) - \alpha_i| < \epsilon \ \forall i>n(\epsilon)$ and all $\bx$. It follows that $\{\alpha_i\}$ is a Cauchy sequence, i.e. $|\alpha_j - \alpha_i| < \epsilon \ \forall i,j>n(\epsilon)$ and all $\bx$, i.e.
\[
\bx^+ (\bW_j - \bW_i) \bx < \epsilon \ \forall \bx
\]
from which it follows that $\lambda_1(\bW_j - \bW_i) < \epsilon$ and thus $\|\bW_j - \bW_i\|\rightarrow 0$ in any norm (since all norms are equivalent \cite{Zhang11MatrixTheory}), i.e. $\{\bW_i\}$ is a Cauchy sequence and thus converges \cite{Miller72SymmetricGroups,Hall03LieGroupsAlgebrasRepresentations}, $\bW_i \rightarrow \bW \le a\bI $. Taking Frobenius norm, one obtains element-wise convergence of this matrix sequence.
\end{IEEEproof}
\vspace*{0.5\baselineskip}

Note that this result generalizes to matrices the well-known fact that any scalar increasing and bounded sequence converges.

To proceed further, observe from the definition of $\bsS_{2m}$ that
\begin{align}
\label{eq.M.P5.5}
\min_{\bW_2 \in \bsS_2} C(\bR,\bW_2) \le \min_{\bW_2 \in \bsS_{2m}} C(\bR,\bW_2).
\end{align}
We prove the equality by contradiction. Assume that
\begin{align}
\label{eq.M.P5.6}
\min_{\bW_2 \in \bsS_2} C(\bR,\bW_2) < \min_{\bW_2 \in \bsS_{2m}} C(\bR,\bW_2)
\end{align}
and let $\bW_2^* = \arg\min_{\bW_2 \in \bsS_2} C(\bR,\bW_2)$ be a minimizer over $\bsS_2$. Then, $\bW_2^* \notin \bsS_{2m}$ (due to the strict inequality) so that there exists $\bW_{21} \in \bsS_2$ such that $\bW_{21} \ge \bW_2^*$ (otherwise $\bW_2^*$ were in $\bsS_{2m}$), $\bW_{21} \neq \bW_2^*$, and $C(\bR,\bW_{21}) \le C(\bR,\bW_2^*)$. If $\bW_{21} \in \bsS_{2m}$, we have a contradiction:
\begin{align}
\label{eq.M.P5.7} \notag
C(\bR,\bW_{21}) &\le C(\bR,\bW_2^*) \\ \notag
&< \min_{\bW_2 \in \bsS_{2m}} C(\bR,\bW_2) \\
&\le C(\bR,\bW_{21}).
\end{align}
Assume further that $\bW_{21} \notin \bsS_{2m}$ so that there exists such $\bW_{22} \in \bsS_{2}$ that $\bW_{22} \ge \bW_{21}$, $\bW_{22} \neq \bW_{21}$, and the process is repeated. In this way, we construct a non-decreasing, bounded sequence $\{\bW_2^*, \bW_{21},...,  \bW_{2i},...\}$, which either terminates in a finite number of steps (when some $\bW_{2k} \in \bsS_{2m}$ so we cannot find a greater one) or it continues indefinitely. In the first case, we have a contradiction and thus the assertion is proved.

In the second case, we claim that the sequence will converge to some $\bW \in \bsS_{2m}$. To see this, first observe that this sequence will converge to some $\bW \in \bsS_2$ (due to Lemma \ref{lemma.M.convergence}, since $\bsS_2$ is bounded and closed and thus compact and the sequence is increasing and bounded; the boundedness can be understood in any norm, since all matrix norms are equivalent). Thus, we have to prove that $\bW \in \bsS_{2m}$. To see this, first observe that $\bW \ge \bW_{2i} \ \forall i$ (since the sequence is increasing). If $\bW \notin \bsS_{2m}$, then there exists $\bW^* \in \bsS_2$ such that $\bW^* \ge \bW \ge \bW_{21}$ so it can be taken as a part of the constructed sequence and thus  $\bW$ cannot be its limit - a contradiction. Therefore, $\bW \in \bsS_{2m}$, as claimed. This, however, results in a contradiction to \eqref{eq.M.P5.6} so that \eqref{eq.M.P5.1} holds. To see \eqref{eq.M.P5.2}, take $\max_{\bR}$ in \eqref{eq.M.P5.5}-\eqref{eq.M.P5.7} and apply the same argument. \hfill\IEEEQED

\subsection{Proof of Theorem \ref{thm.M2}}
\label{app:thm.M2}
First, we observe that
\begin{align}
\label{eq.M.T2.4}
C(\bW_{1},\bW_{2},\bR) \ge C(\bW_{1},\epsilon\bI,\bR) \ \forall \bR, \bW_1,
\end{align}
since $\bW_2 \le \epsilon\bI$ (which follows from $|\bW_2|_2 \le \epsilon$) and $|\bI+\bW\bR|$ is monotonically increasing in $\bW$ for any (positive semi-definite) $\bR$. The lower bound is achieved by $\bW_2 = \epsilon\bI$. Therefore,
\begin{align}
\min_{\bW_2} C(\bW_{1},\bW_{2},\bR) = C(\bW_{1},\epsilon\bI,\bR) \ \forall \bR, \bW_1,
\end{align}
and also
\begin{align}
\label{eq.M.T2.6} \notag
C_w &= \min_{\bW_1} \mathop {\max }\limits_{\bR} C(\bW_1, \epsilon\bI, \bR)\\ \notag
&= \min_{\bW_1} \mathop {\max }\limits_{\bR} \ln \frac{|\bI+\bW_1\bR|}{|\bI+\epsilon\bL|}\\ \notag
&{\mathop = \limits^{(a)}} \min_{\bW_1} \mathop {\max }\limits_{\bR} \sum_i \ln \frac{1+\lambda_i(\bW_1)\lambda_i(\bR)}{1+\epsilon\lambda_i(\bR)} \\ \notag
&{\mathop = \limits^{(b)}} \mathop {\max }\limits_{\{\lambda_i\}} \sum_i \ln \frac{1+(\sigma_i(\bH_0)-\epsilon_1)_+^2\lambda_i}{1+\epsilon\lambda_i}\\
&= C(\bW_{1w},\epsilon\bI,\bR^*)
\end{align}
where (a) follows from the inequality
\begin{align}
\label{eq.M.T2.7}
|\bI+\bW_1\bR| \le \prod_i(1+\lambda_i(\bW_1)\lambda_i(\bR))
\end{align}
which follows from \cite[Theorem 3.3.14(c)]{HornJohnson91TopicsMatrixAnalysis} with $f(x)=\ln(1+x)$, where $\lambda_i(\bW_1), \lambda_i(\bR)$ are ordered likewise and the equality is achieved when $\bW_1, \bR$ have the same eigenvectors; (b) follows from the inequality $\sigma_i(\bH_1) \ge (\sigma_i(\bH_0)-\sigma_1(\Delta\bH))_+$ (see e.g. \cite{Zhang11MatrixTheory,HornJohnson91TopicsMatrixAnalysis}) and $\lambda_i(\bW_1) = \sigma_i^2(\bH_1)$ where the equality is achieved by $\bH_{1w}$.

We further observe that the saddle-point property in \eqref{eq.M.T2.1} is equivalent to (see e.g. \cite{Zeidler86NonlinearFunctionalAnalysisI})
\begin{align}
\label{eq.M.T2.8}
C(\bW_{1w}, \epsilon\bI, \bR) {\mathop \le \limits^{(a)}} C(\bW_{1w}, \epsilon\bI, \bR^*) {\mathop \le \limits^{(b)}} C(\bW_{1}, \bW_{2}, \bR^*)
\end{align}
and we prove these inequalities below thus establishing \eqref{eq.M.T2.1}.

Note that (a) follows from \eqref{eq.M.T2.6} (since $\bR^*$ is the optimal covariance for $\bW_{1}=\bW_{1w}, \bW_{2}=\epsilon\bI$). To prove (b), we need the following technical lemma, which is an  extension of well-known singular value inequalities for a sum and a product of two matrices (see e.g. \cite{Zhang11MatrixTheory,HornJohnson91TopicsMatrixAnalysis}):

\begin{lemma}
\label{lemma.M.SV.ineq}
Let ${\bf A}$, ${\bf B}$ and ${\bf C}$ be $n\times m$ and $m\times m$ matrices, and let the right singular vectors of ${\bf A}$ be the same as the left singular vectors of ${\bf C}$ so that their singular value decompositions (SVD) are ${\bf A}={\bf U\Sigma }_a {\bf V}^+$ and ${\bf C}={\bf V\Sigma }_c {\bf W}^+$, where ${\bf U},{\bf V},{\bf W}$ are unitary and ${\bf \Sigma }_a =\diag\{\sigma _{ai} \},{\bf \Sigma }_c =\diag\{\sigma _{ci} \}$ are ``diagonal'' matrices of singular values of ${\bf A}$ and ${\bf C}$. Assume that $\{\sigma _{ai} \}$ and $\{\sigma _{ci}\}$ are in decreasing order.  Then,
\begin{align}
\label{eq.M.T2.9}
\sigma _i (({\bf A}+{\bf B}){\bf C}) \ge (\sigma _i ({\bf A})-\sigma _1 ({\bf B}))_+ \sigma _i ({\bf C})
\end{align}
where $\sigma _i (({\bf A}+{\bf B}){\bf C})$ are also in decreasing order. The equality is achieved by ${\bf B}=-{\bf U\Sigma }_b {\bf V}^+$, where ${\bf \Sigma }_b =\diag\{\min (\sigma _i ({\bf A}),\epsilon )\}$.
\end{lemma}
\begin{IEEEproof}
The proof is based on the variational characterization of singular values, see \cite{Loyka15} for details.
\end{IEEEproof}

Using this lemma, one obtains:
\begin{align}
\label{eq.M.T2.10} \notag
C_w &= C(\bW_{1w},\epsilon\bI,\bR^*)\\ \notag
&{\mathop = \limits^{(a)}} \sum_i \ln \frac{1+(\sigma_i(\bH_0)-\epsilon_1)_+^2\lambda_i^*}{1 +\epsilon\lambda_i^*}\\ \notag
&{\mathop \le \limits^{(b)}} \mathop \sum_i \ln \frac{1+\sigma_i^2(\bH_1\bR^{*1/2})}{1 +\epsilon\lambda_i^*}\\ \notag
&= C(\bW_{1},\epsilon\bI,\bR^*)\\
&{\mathop \le \limits^{(c)}} C(\bW_{1},\bW_2,\bR^*)
\end{align}
where (a) follows from \eqref{eq.M.T2.6}, (b) follows from Lemma \ref{lemma.M.SV.ineq} applied to ${\bf A} = \bH_0, {\bf B} = \Delta\bH, {\bf C} = \bR^{*1/2}$ (and observing, from \eqref{eq.M.T1.3}, that the singular values of $\bH_0$ and $\bR^{*1/2}$ are ordered likewise), where we have used $\lambda_i(\bR)=\sigma_i^2(\bR^{1/2})$, and (c) follows from \eqref{eq.M.T2.4}. This establishes \eqref{eq.M.T2.8} and thus \eqref{eq.M.T2.1}. \hfill\IEEEQED

\subsection{Proof of Theorem \ref{thm.M6.2}}
\label{app:thm.M6.2}

Using the argument similar to that in \eqref{eq.M.r.T1.2}, it follows that
\begin{align}
\ln |\bI+\bW_2\bR^*| &= \sum_{i=1}^{r_2} \ln (1+\lambda_i(\bW_2\bR^*))\notag\\
&\le \sum_{i=1}^{r_1} \ln (1+\lambda_i(\bW_2)\lambda_i(\bR^*))\notag\\
&\le \sum_{i=1}^{r_1} \ln (1+\epsilon^2 \lambda_i(\bR^*))\notag\\
&= \ln |\bI+\bW_{2w}\bR^*|
\end{align}
and
\begin{align}
\ln |\bI+\bW_1\bR^*| &= \sum_{i=1}^{r_1} \ln (1+\lambda_i(\bW_1\bR^*))\notag\\
&{\mathop = \limits^{(a)}} \sum_{i=1}^{r_1} \ln (1+\sigma_i^2((\bH_0+\Delta\bH)\bR^{*\frac{1}{2}}))\notag\\
&{\mathop \ge \limits^{(b)}} \sum_{i=1}^{r_1} \ln (1+(\sigma_i(\bH_0)-\epsilon_1)_+^2\lambda_i(\bR^*))\notag\\
&= \ln |\bI+\bW_{1w}\bR^*|
\end{align}
for any $\bW_1 \in \bsS_1$ and $\bW_2 \in \bsS_2$, where (a) follows from $\lambda_i(\bR)=\sigma_i^2(\bR)$ and (b) follows from the singular value inequalities in Lemma \ref{lemma.M.SV.ineq}. Combining these two chain inequalities, one obtains
\bal
C(\bW_{1w},\bW_{2w}, \bR^*) \le C(\bW_{1},\bW_{2}, \bR^*)
\eal
which establishes the 2nd inequality in \eqref{eq.M.T2.1a}. The 1st inequality follows from the fact that $\bR^*$ is the optimal covariance under $\bW_1=\bW_{1w}$ and $\bW_2=\bW_{2w}$. Since the saddle-point inequalities in \eqref{eq.M.T2.1a} are equivalent to $\max\min = \min\max$ in \eqref{eq.M.T2.1} (see e.g.  \cite{Zeidler86NonlinearFunctionalAnalysisI}), this also establishes the latter claim. \hfill\IEEEQED

\section*{Acknowledgment}
The authors are grateful to P. Mitran for insightful discussions and suggestions.


\begin{IEEEbiographynophoto}
{Rafael F. Schaefer}
(S'08-M'12) received the Dipl.-Ing. degree in electrical engineering and computer science from the Technische Universit\"at Berlin, Berlin, Germany, in 2007, and the Dr.-Ing. degree in electrical engineering from the Technische Universit\"at M\"unchen, Munich, Germany, in 2012. He was a Research and Teaching Assistant with the Heinrich-Hertz-Lehrstuhl f\"ur Mobilkommunikation, Technische Universit\"at Berlin, from 2007 to 2010, and the Lehrstuhl f\"ur Theoretische Informationstechnik, Technische Universit\"at M\"unchen, from 2010 to 2013. He is currently a Post-Doctoral Research Fellow with the Department of Electrical Engineering, Princeton University, Princeton, NJ, USA. He was a recipient of the VDE Johann-Philipp-Reis Prize in 2013. He was one of the exemplary reviewers of the IEEE COMMUNICATION LETTERS in 2013. Currently, he is an associate member of the IEEE Information Forensics and Security Technical Committee. 
\end{IEEEbiographynophoto}

\begin{IEEEbiographynophoto}
{Sergey Loyka}
was born in Minsk, Belarus. He received the Ph.D. degree in
Radio Engineering from the Belorussian State University (BSUIR), Minsk, Belarus
in 1995 and the M.S. degree with honors from Minsk Radioengineering Institute,
Minsk, Belarus in 1992. Since 2001 he has been a faculty member at the
School of Electrical Engineering and Computer Science, University of Ottawa,
Canada. Prior to that, he was a research fellow in the Laboratory of Communications
and Integrated Microelectronics (LACIME) of Ecole de Technologie
Superieure, Montreal, Canada; a senior scientist at the Electromagnetic Compatibility
Laboratory of BSUIR, Belarus; an invited scientist at the Laboratory
of Electromagnetism and Acoustic (LEMA), Swiss Federal Institute of Technology,
Lausanne, Switzerland. His research areas include wireless communications
and networks, MIMO systems and smart antennas, RF system modeling
and simulation, and electromagnetic compatibility, in which he has published
extensively. Dr. Loyka is a technical program committee member of several
IEEE conferences and a reviewer for numerous IEEE periodicals and conferences.
He received a number of awards from the URSI, the IEEE, the Swiss,
Belarus and former USSR governments, and the Soros Foundation.
\end{IEEEbiographynophoto}


\begin{thebibliography}{10}

\bibitem{Shannon49CommunicationTheorySecrecySystems}
C.~E. Shannon, ``{Communication theory of secrecy systems},'' \emph{Bell Syst.
  Tech. J.}, vol.~28, no.~4, pp. 656--715, Oct. 1949.

\bibitem{Wyner75WiretapChannel}
A.~D. Wyner, ``{The wire-tap channel},'' \emph{Bell Syst. Tech. J.}, vol.~54,
  pp. 1355--1387, Oct. 1975.

\bibitem{Liang09InformationTheoreticSecurity}
Y.~Liang, H.~V. Poor, and S.~{Shamai (Shitz)}, ``{Information theoretic
  security},'' \emph{Foundations and Trends in Communications and Information
  Theory}, vol.~5, no. 4-5, pp. 355--580, 2009.

\bibitem{Jorswieck10SecrecyPhysicalLayer}
E.~A. Jorswieck, A.~Wolf, and S.~Gerbracht, ``{Secrecy on the physical layer in
  wireless networks},'' \emph{Trends in Telecommunications Technologies}, pp.
  413--435, Mar. 2010.

\bibitem{Liu10SecuringWirelessCommunications}
R.~Liu and W.~Trappe, Eds., \emph{Securing Wireless Communications at the
  Physical Layer}.\hskip 1em plus 0.5em minus 0.4em\relax Springer, 2010.

\bibitem{Bloch11InformationTheoreticSecrecy}
M.~Bloch and J.~Barros, \emph{Physical-Layer Security: From Information Theory
  to Security Engineering}.\hskip 1em plus 0.5em minus 0.4em\relax Cambridge
  University Press, 2011.

\bibitem{BiglieriG07MIMOWirelessCommunications}
E.~Biglieri, R.~Calderbank, A.~Constantinides, A.~Goldsmith, A.~Paulraj, and
  H.~V. Poor, \emph{MIMO Wireless Communications}.\hskip 1em plus 0.5em minus
  0.4em\relax Cambridge University Press, 2007.

\bibitem{KhistiWornell10MIMOWiretap1}
A.~Khisti and G.~W. Wornell, ``{Secure transmission with multiple antennas I:
  The MISOME wiretap channel},'' \emph{IEEE Trans. Inf. Theory}, vol.~56,
  no.~7, pp. 3088--3104, Jul. 2010.

\bibitem{KhistiWornell10MIMOWiretap2}
------, ``{Secure transmission with multiple antennas--Part II: The MIMOME
  wiretap channel},'' \emph{IEEE Trans. Inf. Theory}, vol.~56, no.~11, pp.
  5515--5532, Nov. 2010.

\bibitem{Oggier11MIMOWiretap}
F.~Oggier and B.~Hassibi, ``{The secrecy capacity of the MIMO wiretap
  channel},'' \emph{IEEE Trans. Inf. Theory}, vol.~57, no.~8, pp. 4961--4972,
  Aug. 2011.

\bibitem{Liu09MIMOWiretapSecrecy}
T.~Liu and S.~{Shamai (Shitz)}, ``{A note on the secrecy capacity of the
  multiple-antenna wiretap channel},'' \emph{IEEE Trans. Inf. Theory}, vol.~55,
  no.~6, pp. 2547--2553, Jun. 2009.

\bibitem{Bustin09MMSEMIMOWiretap}
R.~Bustin, R.~Liu, H.~V. Poor, and S.~{Shamai (Shitz)}, ``{An MMSE approach to
  the secrecy capacity of the MIMO Gaussian wiretap channel},'' in \emph{Proc.
  IEEE Int. Symp. Inf. Theory}, Seoul, Korea, Jun. 2009, pp. 2602--2606.

\bibitem{Loyka12OptimalSignalingMIMOWiretap}
S.~Loyka and C.~D. Charalambous, ``{On optimal signaling over secure MIMO
  channels},'' in \emph{Proc. IEEE Int. Symp. Inf. Theory}, Cambridge, MA, USA,
  Jul. 2012, pp. 443--447.

\bibitem{LiXXMIMOWiretap}
J.~Li and A.~Petropulu, ``{Transmitter optimization for achieving secrecy
  capacity in Gaussian MIMO wiretap channels},'' \emph{submitted to Trans. Inf.
  Theory}, Sep. 2009, available at http://arxiv.org/abs/0909.2622v1.

\bibitem{Blackwell59Compound}
D.~Blackwell, L.~Breiman, and A.~J. Thomasian, ``{The capacity of a class of
  channels},'' \emph{Ann. Math. Stat.}, vol.~30, no.~4, pp. 1229--1241, Dec.
  1959.

\bibitem{Wolfowitz60SimultaneousChannels}
J.~Wolfowitz, ``{Simultaneous channels},'' \emph{Arch. Rational Mech.
  Analysis}, vol.~4, no.~4, pp. 371--386, 1960.

\bibitem{Liang09CompoundWiretapChannels}
Y.~Liang, G.~Kramer, H.~V. Poor, and S.~{Shamai (Shitz)}, ``{Compound wiretap
  channels},'' \emph{EURASIP J. Wireless Commun. Netw.}, vol. Article ID
  142374, pp. 1--13, 2009.

\bibitem{Bjelakovic13CompoundWiretap}
I.~Bjelakovi\'c, H.~Boche, and J.~Sommerfeld, ``{Secrecy results for compound
  wiretap channels},'' \emph{Probl. Inf. Transmission}, vol.~49, no.~1, pp.
  73--98, Mar. 2013.

\bibitem{Ekrem10MIMOCompoundWiretap}
E.~Ekrem and S.~Ulukus, ``{On Gaussian MIMO compound wiretap channels},'' in
  \emph{Proc. Conf. Inf. Sciences and Systems}, Baltimore, MD, USA, Mar. 2010,
  pp. 1--6.

\bibitem{Khisti11InterferenceAlignmentMIMOCompoundWiretap}
A.~Khisti, ``{Interference alignment for the multiantenna compound wiretap
  Channel},'' \emph{IEEE Trans. Inf. Theory}, vol.~57, no.~5, pp. 2976--2993,
  May 2011.

\bibitem{HeYener10MIMOWiretap}
X.~He and A.~Yener, ``{MIMO wiretap channels with arbitrarily varying
  eavesdropper channel states},'' \emph{IEEE Trans. Inf. Theory}, submitted
  2010, available at http://arxiv.org/abs/1007.4801.

\bibitem{Loyka10FiniteSNRDMTLargeMIMO}
S.~Loyka and G.~Levin, ``{Finite-SNR diversity-multiplexing tradeoff via
  asymptotic analysis of large MIMO systems},'' \emph{IEEE Trans. Inf. Theory},
  vol.~56, no.~10, pp. 4781--4792, Oct. 2010.

\bibitem{Schaefer14CompoundBCC}
R.~F. Schaefer and H.~Boche, ``{Robust broadcasting of common and confidential
  messages over compound channels: Strong secrecy and decoding performance},''
  \emph{IEEE Trans. Inf. Forensics Security}, vol.~9, no.~10, pp. 1720--1732,
  Oct. 2014.

\bibitem{Kobayashi09CompoundMIMOBCC}
M.~Kobayashi, Y.~Liang, S.~{Shamai (Shitz)}, and M.~Debbah, ``{On the compound
  MIMO broadcast channels with confidential messages},'' in \emph{Proc. IEEE
  Int. Symp. Inf. Theory}, Seoul, Korea, Jun. 2009, pp. 1283--1287.

\bibitem{Wolf15WorstCaseSecrecyRatesMIMOME}
A.~Wolf, E.~A. Jorswieck, and C.~R. Janda, ``{Worst-case secrecy rates in
  MIMOME systems under input and state constraints},'' in \emph{Proc. IEEE Int.
  Workshop Inf. Forensics and Security}, Rome, Italy, Nov. 2015.

\bibitem{Mitran06CompoundSideInformation}
P.~Mitran, N.~Devroye, and V.~Tarokh, ``{On compound channels with side
  information at the transmitter},'' \emph{IEEE Trans. Inf. Theory}, vol.~52,
  no.~4, pp. 1745--1755, Apr. 2006.

\bibitem{Csiszar96SecrecyCapacity}
I.~Csisz\'ar, ``{Almost independence and secrecy capacity},'' \emph{Probl.
  Pered. Inform.}, vol.~32, no.~1, pp. 48--57, 1996.

\bibitem{Maurer00WeakToStrongSecrecy}
U.~M. Maurer and S.~Wolf, ``{Information-theoretic key agreement: From weak to
  strong secrecy for free},'' in \emph{EUROCRYPT 2000, Lecture Notes in
  Computer Science}.\hskip 1em plus 0.5em minus 0.4em\relax Springer-Verlag,
  May 2000, vol. 1807, pp. 351--368.

\bibitem{BocheSchaeferPoorXXContinuity}
H.~Boche, R.~F. Schaefer, and H.~V. Poor, ``{On the continuity of the secrecy
  capacity of compound and arbitrarily varying wiretap channels},''
  \emph{IEEE Trans. Inf. Forensics Security}, accepted for publication.

\bibitem{Leung09ContinuityQuantumCapacities}
D.~Leung and G.~Smith, ``{Continuity of quantum channel capacities},''
  \emph{Commun. Math. Phys}, vol. 292, no.~1, pp. 201--215, 2009.

\bibitem{CsiszarKoerner78BroadcastChannelsConfidentialMessages}
I.~Csisz\'ar and J.~K\"orner, ``{Broadcast channels with confidential
  messages},'' \emph{IEEE Trans. Inf. Theory}, vol.~24, no.~3, pp. 339--348,
  May 1978.

\bibitem{vanDijk97SpecialClassBCC}
M.~{van Dijk}, ``{On a special class of broadcast channels with confidential
  messages},'' \emph{IEEE Trans. Inf. Theory}, vol.~42, no.~2, pp. 712--714,
  Mar. 1997.

\bibitem{Zeidler86NonlinearFunctionalAnalysisI}
E.~Zeidler, \emph{Nonlinear Functional Analysis and Its Applications I:
  Fixed-Point Theorems}.\hskip 1em plus 0.5em minus 0.4em\relax Springer, 1986.

\bibitem{Gray09Probability}
R.~M. Gray, \emph{Probability, Random Processes, and Ergodic Properties},
  2nd~ed.\hskip 1em plus 0.5em minus 0.4em\relax Springer, 2009.

\bibitem{Gray11Entropy}
------, \emph{Entropy and Information Theory}, 2nd~ed.\hskip 1em plus 0.5em
  minus 0.4em\relax Springer, 2011.

\bibitem{ElGamal11NetworkInformationTheory}
A.~{El Gamal} and Y.-H. Kim, \emph{Network Information Theory}.\hskip 1em plus
  0.5em minus 0.4em\relax Cambridge University Press, 2011.

\bibitem{Gallager68InformationTheoryReliableCommunication}
R.~G. Gallager, \emph{Information Theory and Reliable Communication}.\hskip 1em
  plus 0.5em minus 0.4em\relax Wiley \& Sons, 1968.

\bibitem{Loyka12CompoundMIMO}
S.~Loyka and C.~D. Charalambous, ``{On the compound capacity of a class of MIMO
  Channels subject to normed uncertainty},'' \emph{IEEE Trans. Inf. Theory},
  vol.~58, no.~4, pp. 2048--2063, Apr. 2012.

\bibitem{Zhang11MatrixTheory}
F.~Zhang, \emph{Matrix Theory: Basic Results and Techniques}, 2nd~ed.\hskip 1em
  plus 0.5em minus 0.4em\relax Springer, 2011.

\bibitem{Loyka13FurtherResultsSecureMIMO}
S.~Loyka and C.~D. Charalambous, ``{Further results on optimal signaling over
  secure MIMO channels},'' in \emph{Proc. IEEE Int. Symp. Inf. Theory},
  Istanbul, Turkey, Jul. 2013, pp. 2019--2023.

\bibitem{Boyd04ConvexOptimization}
S.~P. Boyd and L.~Vandenberghe, \emph{Convex Optimization}.\hskip 1em plus
  0.5em minus 0.4em\relax Cambridge University Press, 2004.

\bibitem{HornJohnson91TopicsMatrixAnalysis}
R.~A. Horn and C.~R. Johnson, \emph{Topics in Matrix Analysis}.\hskip 1em plus
  0.5em minus 0.4em\relax Cambridge University Press, 1991.

\bibitem{Loyka14RankDeficientMIMOWiretap}
S.~Loyka and C.~D. Charalambous, ``{Rank-deficient solutions for optimal
  signaling over secure MIMO channels},'' in \emph{Proc. IEEE Int. Symp. Inf.
  Theory}, Honolulu, HI, USA, Jun. 2014, pp. 201--205.

\bibitem{Balanis05AntennaTheory}
C.~A. Balanis, \emph{Antenna Theory: Analysis and Design}, 3rd~ed.\hskip 1em
  plus 0.5em minus 0.4em\relax Hoboken, New Jersey: Wiley, 2005.

\bibitem{Miller72SymmetricGroups}
W.~Miller, \emph{Symmetry Groups and Their Applications}.\hskip 1em plus 0.5em
  minus 0.4em\relax Academic Press Inc, 1972.

\bibitem{Hall03LieGroupsAlgebrasRepresentations}
B.~C. Hall, \emph{Lie Groups, Lie Algebras, and Representations: An Elementary
  Introduction}.\hskip 1em plus 0.5em minus 0.4em\relax Springer, 2003.

\bibitem{Loyka15}
S.~Loyka and C.~D. Charalambous, ``{Novel matrix singular value inequalities and
  their applications to uncertain MIMO channels},'' \emph{IEEE Trans. Inf. Theory}, submitted, Mar. 2015.
\end{thebibliography}
\end{document}